\documentclass[fleqn,usenatbib]{mnras}

\usepackage{newtxtext,newtxmath}

\usepackage[T1]{fontenc}

\DeclareRobustCommand{\VAN}[3]{#2}
\let\VANthebibliography\thebibliography
\def\thebibliography{\DeclareRobustCommand{\VAN}[3]{##3}\VANthebibliography}

\usepackage{amsmath}	
\usepackage{comment}
\usepackage{graphicx}	
\usepackage[version=4]{mhchem}
\usepackage{threeparttable} 
\usepackage{xspace}

\begingroup
\catcode`\_=\active
\gdef_#1{\ensuremath{\sb{\mathrm{#1}}}}
\endgroup
\mathcode`\_=\string"8000
\catcode`\_=12

\newcommand{\hm}{H$_2$\xspace} 
\newcommand{\hii}{H~II\xspace} 

\newcommand{\intj}{{\rm J}} 

\newcommand{\npop}{n_{pop3}\xspace} 

\newcommand{\hcc}{H~cm$^{-3}$\xspace} 
\newcommand{\mpc}{{\rm Mpc}} 
\newcommand{\mpch}{{\rm Mpc}/h} 
\newcommand{\msun}{{\rm M}_{\odot}} 

\newcommand{\dd}{\mathrm{d}}

\newcommand{\colossus}{{\sc colossus}\xspace} 
\newcommand{\music}{{\sc music}\xspace} 
\newcommand{\ramses}{{\sc ramses}\xspace} 
\newcommand{\ramsesrt}{{\sc ramses-rt}\xspace} 
\newcommand{\rockstar}{{\sc rockstar}\xspace} 

\newcommand{\hide}[1]{}

\title[First stars in an X-ray background]{Population~III star formation in an X-ray background: V. 
Environmental dependence and halo occupation probability}

\author[J. Park and M. Ricotti]{
Jongwon Park$^{1}$\thanks{E-mail: jwpark5064@yonsei.ac.kr}
and Massimo Ricotti$^{2}$
\\
$^{1}$Department of Astronomy, Yonsei University, Seoul 03722, Republic of Korea\\
$^{2}$Department of Astronomy, University of Maryland, College Park, MD 20742, USA\\
}

\date{Accepted XXX. Received YYY; in original form ZZZ}

\pubyear{\the\year{}}

\begin{document}
\label{firstpage}
\pagerange{\pageref{firstpage}--\pageref{lastpage}}
\maketitle

\begin{abstract}
An X-ray background in the early Universe enhances molecular hydrogen formation, the main coolant of primordial gas, thereby lowering the threshold for Pop~III star formation. Continuing our series on X-ray impacts on Pop~III star formation, we investigate how a soft X-ray background promotes Pop~III star formation using cosmological zoom-in simulations of ten cosmic volumes spanning a range of halo number densities. Each volume is irradiated by the Lyman-Werner (LW) \hm dissociating background and a weak ($\intj_{21} \sim 10^{-5}$), soft ($E \sim 0.2$--$2.0$~keV) X-ray background produced by pair-instability SNe (PISNe) from Pop~III stars and calculated self-consistently as described in a companion paper. We also compare the same simulations with and without X-rays to isolate the X-ray effect. The background promotes Pop~III star formation in two ways: (1) by reducing the mean host halo mass by a factor of $\sim 2$--$3$, and (2) by enabling Pop~III star formation in haloes that would otherwise remain sterile, thereby increasing the halo occupation fraction. The resulting gain in Pop~III number density is largest in underdense regions (a factor of $\approx 3$ on average, reaching up to 7). In the most extreme case, Pop~II stars form only in the presence of X-rays and the gas-phase metallicity rises by an order of magnitude, suggesting that dwarf galaxies in underdense regions may be significantly influenced by an early X-ray background. We also provide fitting functions for the halo occupation probability of Pop~III stars as a function of redshift for both X-ray and LW-only simulations, which can serve as inputs for semi-analytic models.
\end{abstract}





\section{Introduction}

The first stars, Population~III (Pop~III) stars, play a central role in early galaxy formation. They synthesise the first heavy elements \citep{greif_simulations_2011, wise_birth_2012, safranek-shrader_star_2014, chiaki_metal-poor_2018, abe_formation_2021} and regulate the formation of the metal-enriched second-generation (Pop~II) stars (\citealp*{ricotti_fate_2002a, ricotti_fate_2002b}; \citealp{wise_resolving_2008, jeon_recovery_2014}), the building blocks of the first galaxies. As the \textit{James Webb Space Telescope} (\textit{JWST}) pushes the cosmic frontier to unprecedented redshifts \citep{bouwens_newly_2019, finkelstein_census_2022, finkelstein_ceers_2023, curtis-lake_spectroscopic_2023, robertson_earliest_2024, carniani_spectroscopic_2024}, the role of Pop~III stars in early galaxy assembly has come into sharper focus. Massive Pop~III stars \citep[$M \sim 140-260~\msun$,][]{abel_formation_2002,bromm_formation_2002} are expected to end their lives as energetic hypernovae or pair-instability supernovae \citep[PISNe,][]{heger_nucleosynthetic_2002}, potentially detectable by the \textit{JWST} and the \textit{Nancy Grace Roman Space Telescope} \citep{whalen_finding_2014}. These prospective observations make theoretical studies of Pop~III star formation particularly timely.

Pop~III star properties have been studied extensively over the past few decades. They are thought to be massive ($M \sim 100~\msun$), as their formation relies on the relatively inefficient \hm cooling \citep{abel_formation_2002, bromm_formation_2002}. Simulations further suggest that they form in binaries or small groups due to protostellar disc fragmentation \citep{stacy_first_2010,clark_formation_2011,clark_gravitational_2011,susa_mass_2014}, with member stars either migrating inward to merge \citep{turk_formation_2009, greif_formation_2012, hirano_formation_2017, chon_forming_2019} or outward to form wide binaries \citep{greif_formation_2012, sugimura_birth_2020, sugimura_formation_2023, park_origin_2024}. Here we focus on a complementary aspect: the number density of Pop~III stars ($\npop$, the number of Pop~III stars per $1~\mpch$ cube)\footnote{For simplicity, we assume one Pop~III star forms per halo, following \citet{ricotti_x-ray_2016}, so the number density of Pop~III stars and Pop~III-forming haloes are identical. We use the terms `Pop~III density' and `Pop~III halo density' interchangeably.}\footnote{Throughout the paper, the `number density' of Pop~III stars/haloes is denoted by $\npop$, while the `number' of stars is denoted by $N_{pop3}$.}, which is tightly linked to the conditions for Pop~III star formation.

The condition for Pop~III star formation is commonly discussed in terms of the critical mass — the minimum dark matter (DM) minihalo mass ($M \sim 10^6~\msun$) above which Pop~III stars can form. However, Pop~III-forming haloes exhibit large scatter in mass (\citealp{hirano_one_2014,hirano_primordial_2015,schauer_influence_2019,skinner_cradles_2020,schauer_influence_2021,kulkarni_critical_2021}; \citealp*{park_population_2021a}),\footnote{We also refer the reader to \citet{nebrin_starbursts_2023} for a compilation of critical masses from 22 publications.} leading to several operational definitions in the literature. For instance, \citet{kulkarni_critical_2021} defined the critical mass as the mass bin in which half the haloes contain cold, dense gas, while \citet[][PRS21]{park_population_2021a} used the host halo mass at the time a Pop~III progenitor forms — an approach suited to individual-halo analysis but limited in statistical scope, as only three haloes were studied. \citet[][R16]{ricotti_x-ray_2016} instead defined it as the threshold above which haloes have formed Pop~III stars by a given redshift. Despite these differences, all definitions share the same physical basis: in low-mass haloes, the rates of \hm formation and cooling are insufficient to trigger the collapse of a metal-free gas cloud within a Hubble time. Using this framework, \citet{tegmark_how_1997} estimated the virial mass of the first Pop~III-forming haloes to be $M_{vir} \sim 10^6~\msun$.

Since \hm cooling dominates in metal-free gas, the conditions for Pop~III star formation are sensitive to radiation backgrounds that regulate the \hm fraction. In particular, \hm-dissociating Lyman-Werner (LW\footnote{In Section~\ref{sec:local}, we use LW and FUV (far-UV) interchangeably.}, $11.2~{\rm eV} \leq E \leq 13.6~{\rm eV}$) radiation has been extensively studied: LW photons from Pop~III stars delay and suppress subsequent Pop~III star formation (\citealp*{haiman_radiative_2000}; \citealp*{machacek_simulations_2001}; \citealp*{oshea_population_2008}; \citealp{regan_emergence_2020, incatasciato_modelling_2023}). Ionising radiation, by contrast, can enhance gas-phase \hm formation through
\begin{equation}
    \label{eq:hminus}
    \begin{split}
        &\ce{ H + e^- -> H^- + \gamma }, ~~~~~~~~~~
        &\ce{ H + H^- -> H_2 + e^- },
    \end{split}
\end{equation}
by boosting the electron fraction. Ionising UV\footnote{The term EUV (extreme UV) is used frequently throughout this paper.} ($E \geq 13.6$~eV) from Pop~III stars can provide such positive feedback \citep{ricotti_fate_2002b,susa_secondary_2006,ahn_does_2007,yoshida_early_2007}, but its impact is confined to local scales \citep[$\lesssim 100$~kpc, i.e., the typical size of \hii bubbles,][]{ricotti_fate_2002b,susa_secondary_2006,ahn_does_2007} owing to efficient absorption by the neutral intergalactic medium ($x_{e} \sim 10^{-4}$).

Here we focus on a distinct source of ionising radiation: X-rays. Produced by Pop~III supernovae, high-mass X-ray binaries (HMXBs), or active galactic nuclei (AGNs), X-ray photons penetrate the neutral IGM, travel over large distances ($\gtrsim 10~\mpc$, see fig.~1 of R16), and build up a global background. This background raises the \hm fraction across wide regions of the early Universe, potentially promoting Pop~III star formation on cosmological scales when moderate in intensity.\footnote{In a strong X-ray background, heating dominates and suppresses star formation even when the electron fraction remains elevated as reported by R16 and PRS21.} Such positive X-ray feedback can offset negative LW feedback \citep{haiman_radiative_2000, glover_radiative_2003} or enable \hm cooling in low-mass haloes that would otherwise fail to form stars. 

Although some authors \citep{machacek_effects_2003} pointed out that X-ray positive feedback is mild and therefore does not affect Pop~III star formation significantly, R16 suggested this positive feedback reinforces over time, as illustrated in Fig.~1 of the companion paper \citep[hereafter PR26a]{park_population_2026} and plays an important role. R16 explored this feedback loop and its impact on the critical mass and Pop~III halo number density using semi-analytic models, arguing that PISNe could produce a moderate X-ray background that lowers the critical mass to $M_{halo} \sim 10^5~\msun$ and raises the Pop~III number density to $\sim 400$ per $(\mpch)^3$ — more than an order of magnitude above the weak-X-ray case. PRS21 qualitatively confirmed this trend with zoom-in simulations of individual haloes, finding that host halo masses can decrease by up to a factor of $\sim 10$ in moderate or strong X-ray backgrounds. Building on these studies, we further investigate X-ray feedback using cosmological simulations.

\begin{figure}
    \centering
	\includegraphics[width=0.48\textwidth]{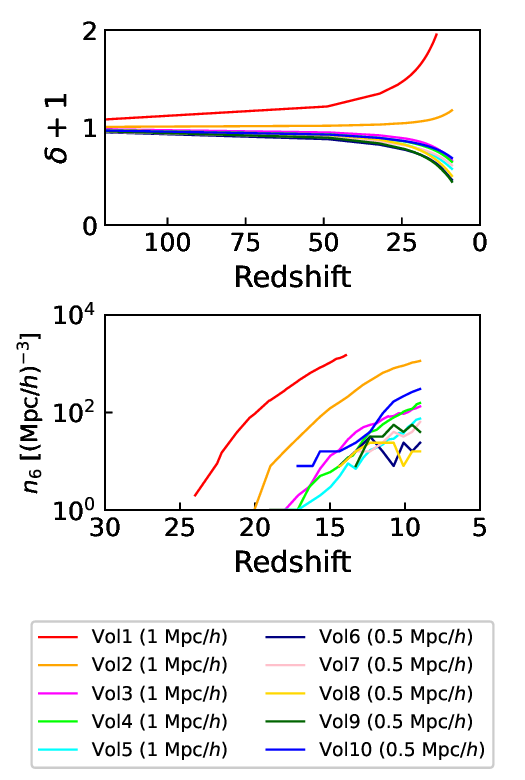}
    \caption{\textbf{Top:} Overdensity as a function of redshift for the 10 simulations used in this study. Volume~1 corresponds to an overdense region, and Volume~2 is a mean-density region. The remaining volumes correspond to underdense regions. \textbf{Bottom:} Number density of haloes more massive than $10^6~\msun$ ($n_6 = n(M > 10^6~\msun)$) as a function of redshift.}
    \label{fig:delta}
\end{figure}

A key step in understanding X-ray feedback is estimating the X-ray background and evaluating its impact on primordial gas in a cosmological context. In PR26a, we extend the approaches of earlier pioneering studies \citep{haardt_radiative_2012,jeon_radiative_2014,xu_heating_2014,ahn_spatially_2015,hummel_first_2015,xu_x-ray_2016,madau_radiation_2017} by introducing an `on-the-fly' method to compute global X-ray and LW backgrounds. This method self-consistently captures the interplay between Pop~III stars and radiation, including the reinforcing feedback loop. In PR26a we focused on describing this methodology and examining the global effects of the X-ray background on the IGM before reionisation. We found that a moderate LW background [$\intj_{21} \sim 10^{-2}$, where $\intj_{21} \equiv J_\nu/(10^{-21}~{\rm erg}~{\rm s}^{-1}~{\rm cm}^{-2}~{\rm Hz}^{-1}~{\rm sr}^{-1})$] and a weak X-ray background ($\intj_{21} \sim 10^{-5}$) are established and persist until $z \sim 12$. After the onset of Pop~II star formation, the LW background rapidly intensifies ($\intj_{21} \gtrsim 10^{1}$), suppressing Pop~III star formation at $z \sim 12$. The X-ray background, though weak, partially offsets the strong negative LW feedback, sustaining Pop~III star formation and increasing $\npop$ across environments. In this paper, we further investigate positive X-ray feedback, including its effects on the critical mass, local UV feedback, and environmental dependence.

The paper is organised as follows. Section~\ref{sec:method} summarises the cosmological simulations and the on-the-fly radiation background calculation. Results are presented in subsequent sections: Section~\ref{sec:mhalo} examines the masses and number densities of Pop~III-hosting haloes across environments, while Sections~\ref{sec:local} investigate the roles of local UV feedback and halo growth rates in driving the environmental differences identified in Section~\ref{sec:mhalo}. Discussion and summary are given in Sections~\ref{sec:disc} and \ref{sec:summary}, respectively.


\section{Methods}
\label{sec:method}

We performed cosmological zoom-in simulations using the radiative hydrodynamics code \ramsesrt \citep{teyssier_cosmological_2002,rosdahl_ramses-rt_2013}. Initial conditions were generated with \music \citep{hahn_multi-scale_2011}, and haloes identified with the \rockstar halo finder \citep{behroozi_rockstar_2013}. The version of \ramsesrt used here includes a primordial chemical network and cooling functions, and has been used in our previous studies of the critical mass (PRS21) and Pop~III properties \citep*{park_population_2021b,park_population_2023,park_origin_2024}. We refer the reader to PRS21 and PR26a for technical details; below we summarise the key aspects of our methodology.

\begin{figure}
    \centering
	\includegraphics[width=0.48\textwidth]{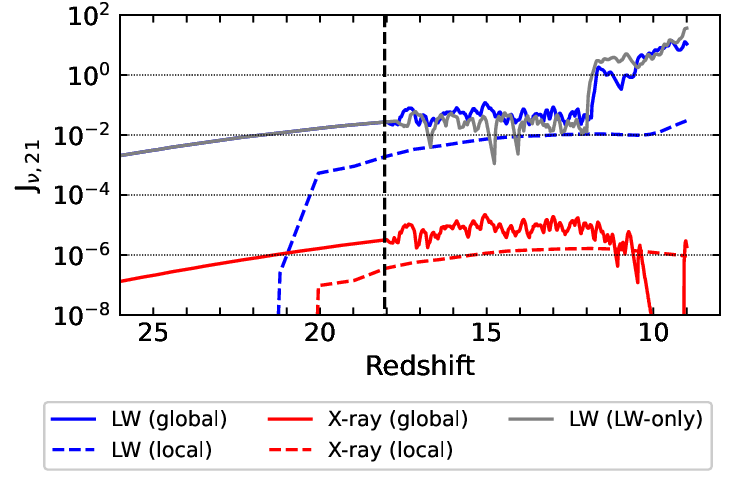}
    \caption{Global radiation backgrounds calculated on-the-fly from the simulations of Volume~2. Solid lines show the intensities (in units of $\intj_{21}$) of the X-ray ($E = 0.36$~keV; red) and LW (blue) backgrounds as a function of redshift in the V2X simulation (Volume~2 with X-rays+LW). Note that $J_{X} \equiv \nu J_\nu$ often reported in the literature is related to $J_{\nu, 21}$ by the relationship $J_{X} = 0.87\times 10^{-4} (E/0.36~{\rm keV})J_{\nu,21}$. The LW background from V2LW (Volume~2 with LW-only) is shown in grey. Dashed lines denote the corresponding local backgrounds for comparison. The vertical dashed line show the redshift where the calculation of the backgrounds switches from the analytic approximation to on-the-fly calculation (see PR26a).}
    \label{fig:spec_total}
\end{figure}

\subsection{Radiation Background Estimation}
\label{sec:method1}

To investigate X-ray effects across different environments, we carried out cosmological zoom-in simulations of regions with varying halo number densities. Ten sub-volumes, each with sizes of $1.2~\mpch$ or $0.6~\mpch$, were selected from two DM-only simulations with box sizes of $8~\mpch$ or $4~\mpch$, respectively. Among these, Volume~1 represents an overdense region with a halo number density higher than the cosmic mean, whereas Volume~2 corresponds to an average region. The remaining sub-volumes (Volumes~3-10) are underdense, with halo densities lower than the average.

\begin{table}
    \caption{Summary of the simulations}
    \centering
    \footnotesize
    \begin{threeparttable}
        \begin{tabular}{ | l | c | c | c | c | c | }
		\hline
            ~~Volume & $L_{box}$\tnote{a} & $L_{zoom}$\tnote{b} & $L_{tar}$\tnote{c} & 
            $1+\delta$\tnote{d} \\
            
            \hline
            Volume 1 & $8$ & $1.2$ & $1.0$ & 1.99   \\
            
            \hline
            Volume 2 & $8$ & $1.2$ & $1.0$ & 1.18 \\
            
            \hline
            Volume 3 & $8$ & $1.2$ & $1.0$ & 0.66 \\
            
            \hline
            Volume 4 & $8$ & $1.2$ & $1.0$ & 0.65 \\
            
            \hline
            Volume 5 & $8$ & $1.2$ & $1.0$ & 0.58 \\
            
            \hline
            Volume 6 & $4$ & $0.6$ & $0.5$ & 0.46 \\
            
            \hline
            Volume 7 & $4$ & $0.6$ & $0.5$ & 0.61 \\
            
            \hline
            Volume 8 & $4$ & $0.6$ & $0.5$ & 0.50 \\
            
            \hline
            Volume 9 & $4$ & $0.6$ & $0.5$ & 0.44 \\
            
            \hline
            Volume 10 & $4$ & $0.6$ & $0.5$ & 0.69 \\
            
            \hline

	\end{tabular}
        \begin{tablenotes}
	        \item[a] Size of the entire box ($\mpch$).
                
                \item[b] Size of the zoom-in region ($\mpch$). This includes the target region and the padding.
                
                \item[c] Size of the target region ($\mpch$).
                
                \item[d] Dark matter overdensity of the target region. $ \rho / \rho_{mean} \equiv 1+\delta$.
        \end{tablenotes}
    \end{threeparttable}
    \label{tab:simulation}
\end{table}

\begin{figure*}
    \centering
	\includegraphics[width=0.95\textwidth]{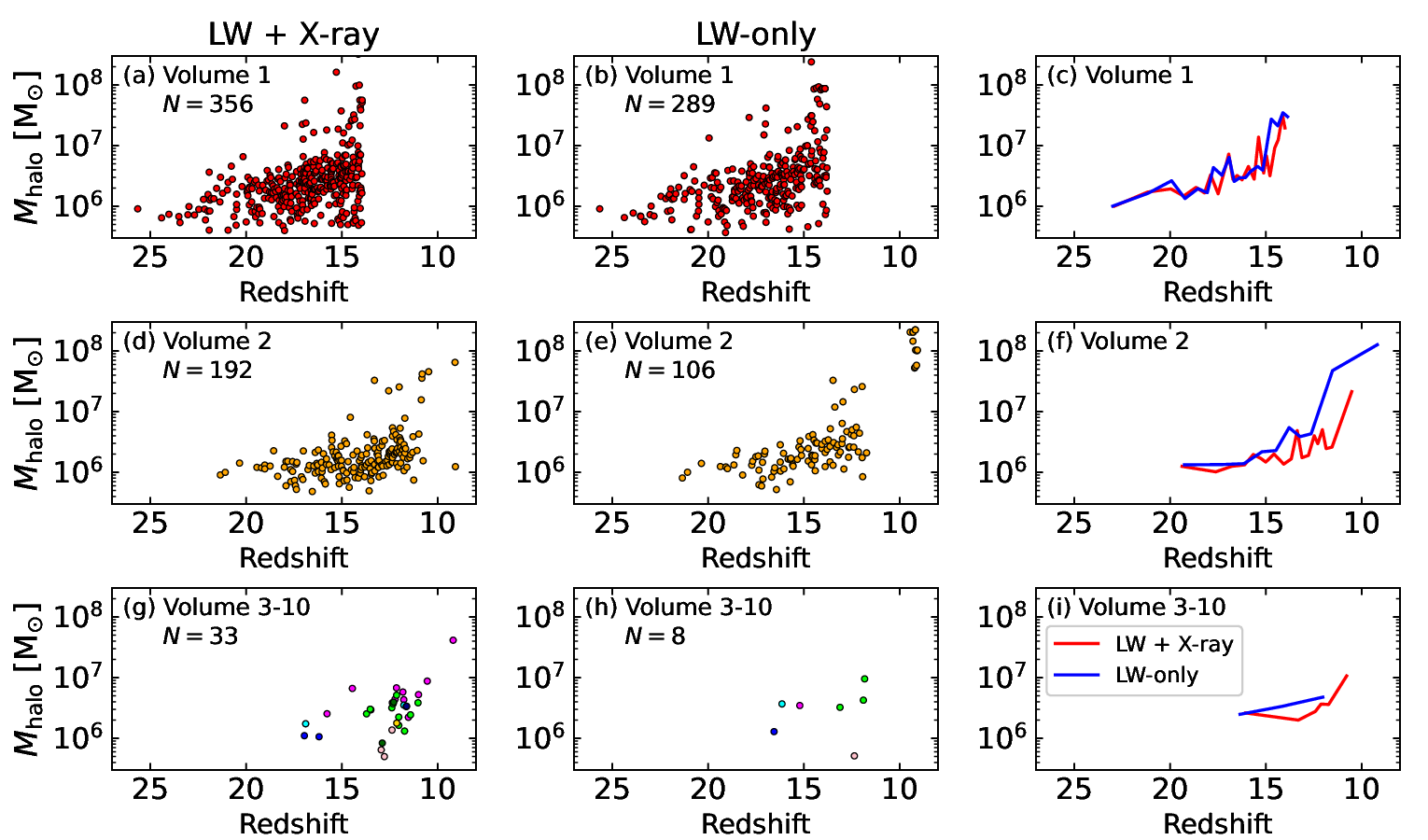}
    \caption{Redshifts of Pop~III star formation and the corresponding masses of their host haloes. Results are shown for Volume~1 (overdense region), Volume~2 (mean-density region), and the remaining volumes (underdense regions), from top to bottom. The left and middle columns show the results of the LW + X-ray (left) and LW-only (middle) simulations, respectively. In each panel the legend shows the number of Pop~III hosting haloes. The right panels present the average halo masses over several redshift intervals for the two simulations. At a given redshift, Pop~III-hosting haloes span a wide mass range with no significant difference of the minimum and maximum masses between the X-ray and LW-only cases. However, as indicated by the differing number of points, Pop~III-hosting halo counts increase in the presence of the X-ray background.}
    \label{fig:mhz}
\end{figure*}

The central $1.2~\mpch$ or $0.6~\mpch$ are defined as the zoom-in regions, where the grids in these regions are allowed to be refined and reach spatial resolutions of $3.8$ or $1.9$ comoving pc, respectively. Within each zoom-in region, we further defined a `target region' of size $1.0~\mpch$ or $0.5~\mpch$, which is less affected by numerical artefacts that might arise near the zoom-in boundaries. The results presented in this paper are derived from these target regions.

To assess the impact of X-ray backgrounds, we performed two simulations for each sub-volume: one including only a LW background (e.g., V2LW) and another incorporating X-ray effects in addition to the LW background (e.g., V2X), resulting in twenty simulations in total. Table~\ref{tab:simulation} presents a summary of the key parameters for each volume; a complete version is provided in PR26a. We also refer the reader to fig.~3 of PR26a for density maps and halo mass functions of these regions. The DM overdensities and halo number densities ($M_{halo} > 10^6~\msun$) are shown in Fig.~\ref{fig:delta}.

We first carried out simulations of Volume~2 (V2LW and V2X) to compute the global LW and X-ray backgrounds on-the-fly. Volume~2 was selected because its halo number density is close to the cosmic mean (see fig.~3 of PR26a), making its star formation history representative of the Universe as a whole. We refer the reader to Appendix~\ref{sec:vol2} for a comparison between Volume~2 and the PS fits. The resulting global backgrounds are shown in Fig.~\ref{fig:spec_total}. The X-ray intensity is $3$--$4$ orders of magnitude lower than the LW intensity until $z \sim 12$. With the onset of Pop~II star formation, the LW intensity rises sharply at $z \sim 12$, while the X-ray intensity declines as Pop~III supernovae (SNe) are the sole X-ray sources. The dashed lines in the figure show the mean local LW and X-ray background intensities,
\begin{equation}
    \intj_{loc,21} = 1.6 \times 10^{-65} \left( \frac{n_{ph}}{1~{\rm Mpc}^{-3}} \right) \times \left( \frac{1+z}{31} \right)^3,
\end{equation}
where $n_{ph}$ is the comoving number density of LW or X-ray photons in Mpc$^{-3}$, estimated by dividing the total photon count in the corresponding bin (see Table~2 of PR26a) by the comoving volume of the target region. This formula follows equation~(18) of \citet{trenti_formation_2009}. The computed backgrounds were tabulated and applied to the remaining simulations. We refer the reader to Fig.~2 of PR26a for a schematic overview of our background estimation method.

\subsection{Star Formation and Feedback}
\label{sec:method2}

\textbf{Pop~III star: } We used the built-in clump finder \citep{bleuler_phew_2015} to identify gas clumps and place $100~\msun$ Pop~III star particles at their peak locations if they satisfy the density ($n_{H} > 10^4$~\hcc) and metallicity ($Z < 10^{-4}$~Z$_{\odot} = 2 \times 10^{-6}$) criteria. Once formed, Pop~III stars emit UV radiation during their main-sequence lifetime ($\sim 2$~Myr) and explode as a PISNe with total energy $E_{PISN} = 10^{53}$~ergs. Each supernova ejects $20~\msun$ of metals and the total energy emitted in soft X-ray band is  $6 \times 10^{50}$~ergs (see R16).

\textbf{Pop~II star: } Pop~II stars form when gas cells exceed the density threshold for star formation ($n_{H} > 10^4$~\hcc) and the critical metallicity ($Z > 10^{-4}$~Z$_{\odot}$). The star formation rate of Pop~II stars follows the Schmidt law \citep{schmidt_rate_1959} with a fixed star formation efficiency ($\epsilon_{SF}=0.1$). Since each Pop~III SN ejects $20~\msun$ of metals, the mean halo metallicity remains above the critical value if the halo retains less than $10^7~\msun$ of gas. As a result, most haloes ($M_{halo} \lesssim 10^8~\msun$) in our simulation transition from Pop~III to Pop~II star formation after the first Pop~III SN event. Although here we assume the equivalent of one PISN per Pop~III star formation episode, Pop~III typically form in multiple star system \citep{park_population_2021b}, hence can produce multiple SN explosions at once\footnote{Our fiducial model is therefore equivalent to a multiple system of $N$ Pop~III stars that either all explode with energy $E_{PISN}/N$, or only one explodes as a PISN while the others fail to explode.}.

\section{Results}
\subsection{Mass of Pop~III-Hosting Haloes and Their Number Densities}
\label{sec:mhalo}

The dark matter halo mass is a key parameter for Pop~III star formation in minihaloes, since the virial temperature ($T \propto M_{vir}^{2/3}$) controls the rates of gas-phase \hm formation (Eq.~\ref{eq:hminus}) and \hm cooling \citep{tegmark_how_1997}. In low-mass haloes, these rates are too low to initiate gas cooling, condensation, and star formation. A weak or moderate X-ray background can boost the electron fraction, enhance \hm formation and cooling, and thereby enable Pop~III star formation in haloes that would otherwise remain sterile — a positive feedback confirmed by PRS21, who showed that X-rays reduce the critical mass and increase $\npop$ at a given redshift.

\begin{figure}
    \centering
	\includegraphics[width=0.48\textwidth]{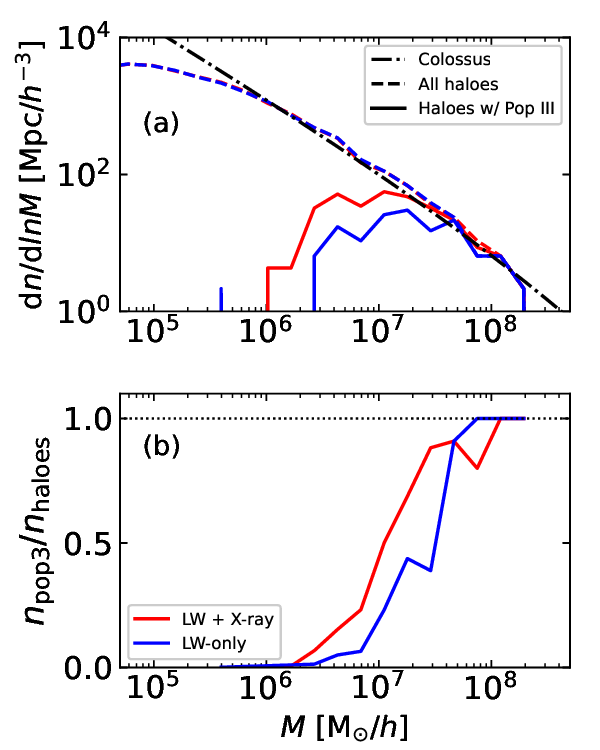}
    \caption{\textbf{Panel~a:} Halo mass functions at $z=9$. Dashed lines show the mass functions of all haloes in Volume~2; solid lines indicate Pop~III-hosting haloes. The Press-Schechter (PS) mass function computed with \colossus \citep{diemer_colossus_2018} is shown as a dot-dashed line. The mass functions of V2X and V2LW agree well with the PS fit at $z = 9$. Colours indicate the presence of an X-ray background (see legend in Panel~b). Note that Fig.~\ref{fig:mhz} shows halo masses at the epoch of Pop~III star formation, whereas this figure shows masses at $z = 9$ regardless of when stars formed. \textbf{Panel~b:} Occupation fraction, defined as the ratio of Pop~III-forming haloes to all haloes in each mass bin.}
    \label{fig:mf}
\end{figure}

We revisit this positive X-ray feedback in a cosmological context, focusing on overall trends across the halo mass function and the resulting evolution of $\npop$. Fig.~\ref{fig:mhz} summarises our results, showing the redshifts and host halo masses at the time of Pop~III star formation for the X-ray+LW (left column), LW-only (middle column), and mean halo masses as a function of redshift (right column). The three rows correspond, from top to bottom, to Volume~1 (overdense), Volume~2 (mean density), and Volumes~3--10 (underdense).

The key findings are as follows.
\begin{enumerate}
    \item Pop~III-forming halo masses show large scatter at any given redshift, with the mean and scatter both evolving with time. The minimum mass remains roughly constant at $5\times 10^5~\msun$, while the high-mass end grows from $10^6~\msun$ at $z\sim 25$ to $10^8~\msun$ at $z\sim 12$, tracking the general growth of dark matter haloes.

    \item Despite similar mass distributions and mean halo masses, the X-ray simulations produce more Pop~III stars than the LW-only runs (compare left and middle panels of Fig.~\ref{fig:mhz}), indicating that positive X-ray feedback acts primarily by enabling Pop~III star formation in haloes that would otherwise remain inactive.
    
    \item X-rays do not produce a systematic reduction in the critical mass as might be naively expected. In Volume~1, the mean halo masses in the X-ray and LW-only runs are similar. In other volumes, the mean halo mass is lower in the X-ray case by at most a factor of a few. In Volume~2, the difference becomes more pronounced below $z \sim 12$, when the LW background strengthens.
\end{enumerate}
These results reflect the combined effects of earlier Pop~III star collapse in growing haloes and increased halo occupancy.

X-rays can trigger Pop~III star formation earlier, modestly reducing host halo masses. PRS21 showed that for weak X-ray backgrounds ($\intj_{21} \lesssim 10^{-5}$), mass reduction is limited to a factor of $\sim 2$ (their figs.~5 and 6). The X-ray background estimated in PR26a is comparably weak, so the two results are consistent. X-rays also enable Pop~III star formation in haloes that would otherwise remain sterile under LW feedback — as seen in fig.~6 of PRS21, where low-mass haloes (Haloes~2 and 3, $M \lesssim 10^7~\msun$ by $z = 9$) fail to form Pop~III stars under moderate or strong LW backgrounds but successfully host them once weak X-rays are introduced. This is reflected in our simulations as an increased number of Pop~III-forming haloes in the X-ray runs (compare left and middle panels of Fig.~\ref{fig:mhz}).

To illustrate the change in Pop~III halo occupancy, we compare the halo mass functions in V2X and V2LW at $z=9$ (Fig.~\ref{fig:mf}). In the X-ray run (red lines), the mass function of Pop~III-hosting haloes is higher than its LW-only counterpart. The occupation fractions\footnote{Defined as the ratio of Pop~III haloes to all haloes ($\npop/n_{haloes}$) in each mass bin.} (Panel~b) increase across most halo masses with X-rays, confirming positive feedback. The trend reverses at the highest masses ($M > 5 \times 10^7~\msun$), but the small number of haloes makes this difference statistically insignificant.

Although the critical mass is a widely used concept, Pop~III-forming haloes span a broad mass range with no sharp threshold (Fig.~\ref{fig:mhz}; Panel~b of Fig.~\ref{fig:mf}). We therefore adopt the definition of \citet{kulkarni_critical_2021}, who set the critical mass as the mass at which the occupation fraction reaches 50\%. With the Pop~III-forming halo mass function shifted to lower masses in the X-ray case, the corresponding critical mass at $z\sim 9$ decreases by a factor of $\sim 3$ ($\sim 10^7~\msun$ vs.\ $3 \times 10^7~\msun$).

\begin{figure}
    \centering
	\includegraphics[width=0.48\textwidth]{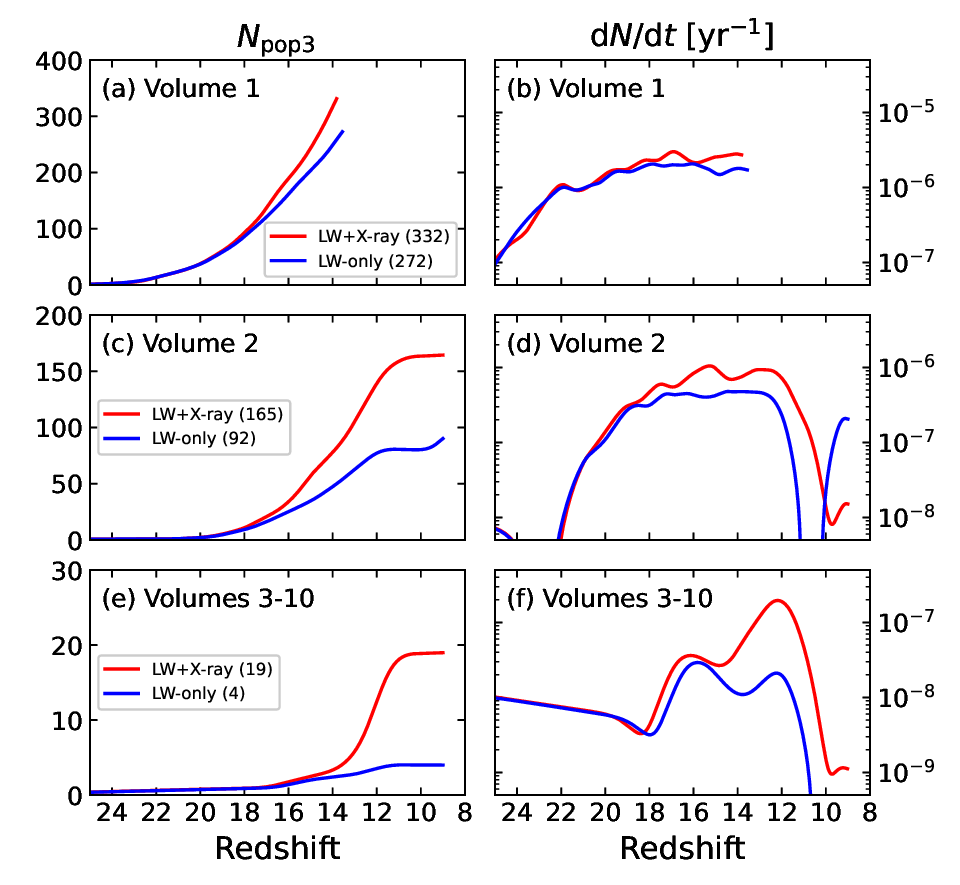}
    \caption{\textbf{Left:} Cumulative number of Pop~III stars in the X-ray (red lines) and LW-only (blue lines) simulations. The final total count of Pop~III stars is indicated in the legend of each panel. \textbf{Right:} Pop~III star formation rates (SFRs) in number per year. Since $M_{pop3} = 100~\msun$, multiplying by the Pop~III mass yields the SFR in $\msun/$yr. From top to bottom, we show the results of Volume~1 and Volume~2, and low-density regions in total (Volumes~3-10), respectively.}
    \label{fig:npop3}
\end{figure}

In summary, X-rays promote Pop~III star formation by enhancing the occupation fraction and hence the number density of Pop~III-forming haloes, but this effect appears to be strongly dependent on the cosmic environment, as discussed in the next section. 

\subsection{Environmental Effects}
\label{sec:local}

Fig.~\ref{fig:npop3} shows the cumulative number of Pop~III stars (left panels) and its time derivative (right panels) in Volume~1 (overdense), Volume~2 (mean density), and Volumes~3--10 (underdense) as a function of redshift. Three key features stand out.
\begin{enumerate}
    \item In all environments, X-ray simulations (red lines) consistently produce more Pop~III stars than LW-only runs (blue lines). The gap grows with decreasing redshift and becomes most pronounced below $z\sim 12$, when the LW background intensifies due to Pop~II star formation (Fig.~\ref{fig:spec_total}).
    
    \item The X-ray enhancement is stronger in underdense regions (Panels~e and f). In Volume~1 the X-ray background increases $N_{pop3}$ by only a factor of $\approx 1.2$; in Volume~2 by $\approx 1.8$. In the low-density volumes (3--10) combined, the total number of Pop~III stars rises by a factor of $\sim 4$ with X-rays.

    \item The enhancement is smaller than predicted by our earlier analytic (R16) and numerical (PRS21) works, which anticipated nearly a factor of 10 increase at $z\sim15$ when including X-rays from PISNe.

    \item In the mean-density region (Volume~2), the Pop~III star formation rate remains $\dd N / \dd t \sim 10^{-6}$~yr$^{-1}$ between $z \sim 20$ and $z \sim 10$. This corresponds to a star formation rate density of $\sim 10^{-4}~\msun$~yr$^{-1}$~(Mpc/$h$)$^{-1}$, consistent with the values reported by previous studies \citep{skinner_cradles_2020,wells_connecting_2022,hegde_self-consistent_2023,hegde_efficient_2025}, although Pop~III star formation continues down to $z \sim 6$ in \citet{hegde_self-consistent_2023} and \citet{hegde_efficient_2025}.
\end{enumerate}

\begin{figure}
    \centering
	\includegraphics[width=0.48\textwidth]{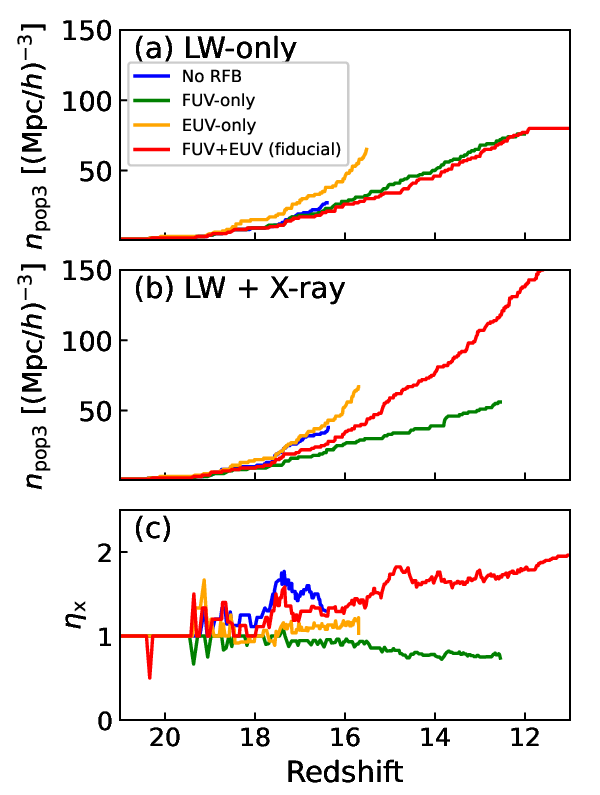}
    \caption{Numerical experiments for local stellar radiative feedback effects. We compare four simulations of Volume~2 (mean-density region) with different types of local radiative feedback: no radiative feedback (No RFB, blue lines), FUV-only (green lines), EUV-only (orange lines), and FUV + EUV (red lines, fiducial). \textbf{Panel~a:} Number densities of Pop~III stars ($\npop$) as functions of redshift without the X-ray background. The No RFB, FUV-only, and fiducial runs show similar results, whereas the EUV-only run yields a higher $\npop$ than the others. \textbf{Panel~b:} Same as Panel~a but with the X-ray background. Unlike in the LW-only case, the FUV-only run shows a lower $\npop$ than the others, while Pop~III formation is promoted in the No RFB and EUV-only runs. \textbf{Panel~c:} Enhancement of Pop~III star formation by X-rays ($\eta_{\mathrm{X}} = n_{pop3,X} / n_{pop3,LW}$). Positive X-ray feedback is less effective in the EUV-only run than in the fiducial case. Pop~III star formation is slightly suppressed by the X-ray in the FUV-only case, suggesting that negative feedback becomes more significant. No RFB run shows only a minor enhancement in the Pop~III number density.}
    \label{fig:local}
\end{figure}

We hypothesise that the environmental dependence stems from either local UV feedback or from environmentally dependent halo growth rates, and investigate both in the following subsections.

\begin{figure*}
    \centering
	\includegraphics[width=0.95\textwidth]{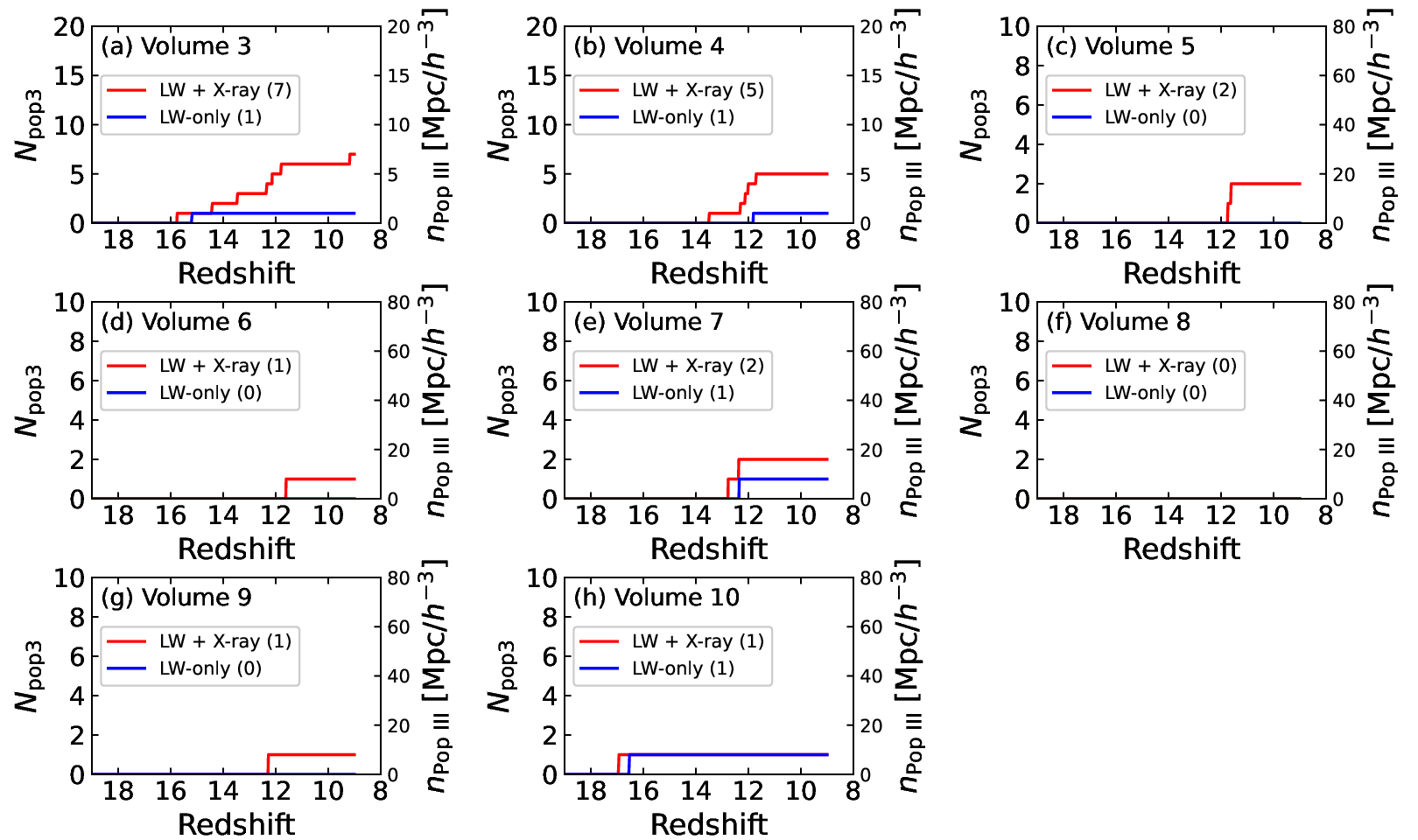}
    \caption{Number of Pop~III stars in each low-density volume (Volumes~3--10). The format follows Fig.~\ref{fig:npop3}. The right-hand $y$-axis gives the number density ($N_{pop3}$ divided by the target region volume). No Pop~III-forming haloes are found in Volume~8 (Panel~f). The strongest X-ray enhancement occurs in Volume~3 (Panel~a, factor of 7); on average, X-rays boost the Pop~III halo number density by a factor of 3.}
    \label{fig:npop3_low}
\end{figure*}

\subsubsection{Local UV Feedback}

UV radiation from Pop~III and Pop~II stars is known to influence neighbouring Pop~III star formation. LW (FUV) radiation photodissociates \hm, suppressing or delaying Pop~III star formation in nearby haloes \citep{regan_emergence_2020}, whereas ionising UV (EUV) promotes \hm reformation and reduces the characteristic mass of second-generation Pop~III stars \citep{yoshida_early_2007}. LW radiation can act over large distances \citep{ricotti_fate_2002a}, while EUV is confined to local scales \citep{ricotti_fate_2002b,susa_secondary_2006,ahn_does_2007}. Local UV feedback may therefore play an important role in densely populated regions (Volumes~1 and 2), while underdense regions — where haloes are widely separated — are less affected, allowing X-ray feedback to dominate. Since local UV feedback was neglected in R16 and PRS21, we investigate its combined effects with X-rays through dedicated numerical experiments.

Starting from the fiducial Volume~2 simulations (V2X and V2LW, including both EUV and FUV), we run additional simulations with EUV, FUV, or both selectively disabled to isolate their roles. Fig.~\ref{fig:local} compares the resulting Pop~III number densities with those from the fiducial runs.

Without the X-ray background (Panel~a), the No RFB, FUV-only, and fiducial runs yield similar results. The FUV-only run shows only a minor decline in $\npop$ relative to No RFB, which we attribute to the global LW background dominating over the local FUV contribution (Fig.~\ref{fig:spec_total}), rendering local dissociation largely ineffective. The EUV-only run (orange line), however, shows an increase in Pop~III star counts, highlighting the role of local positive feedback in the absence of X-rays. In the fiducial run, negative FUV and positive EUV feedback cancel, leaving a negligible net effect.

With the global X-ray background (Panel~b), the behaviour changes. The FUV-only run now shows a lower $\npop$ than No RFB, indicating that local FUV feedback counteracts positive X-ray feedback. By contrast, EUV feedback has no further effect once Pop~III star formation is promoted by X-rays. When both FUV and EUV are present, the result falls between these two cases. It is also notable that in the absence of X-rays (Panel~a), the fiducial run does not show a substantial increase over the FUV-only case, while with X-rays (Panel~b) the fiducial run shows a significant increase in $\npop$ with respect to the FUV-only run, suggesting that the X-ray background compensates for the combined negative feedback of global and local LW radiation.

The X-ray enhancement factor ($\eta_{\mathrm{X}} = n_{pop3,X} / n_{pop3,LW}$) is shown in Panel~c for each run. The EUV-only run (orange) gives $\eta_X\sim 1$, indicating limited synergy between EUV and X-ray feedback. In the FUV-only run (green), $\eta_{X}$ falls below unity: without EUV to maintain ionisation, the \hm fraction remains low, and X-ray heating produces net negative feedback. We conclude that EUV alone does not significantly promote Pop~III star formation, but it compensates for LW negative feedback, making positive X-ray feedback more effective. In the fiducial case, X-ray feedback yields a net increase in $\npop$ by up to a factor of two.

Since it appears that the net effect of local UV feedback is small when the X-rays background is included, we therefore turn in Section~\ref{sec:npop3} to investigate the role of halo growth rates as the primary driver.

\subsubsection{Halo Mass Function in Underdense Regions}
\label{sec:npop3}

We now examine how the redshift at which haloes above the minimum mass for Pop~III star formation first appear differs between overdense and underdense regions. As shown in Fig.~\ref{fig:delta}, haloes in overdense regions grow more rapidly, so those massive enough to trigger \hm cooling (e.g., $M \sim 10^6~\msun$) appear at higher redshifts. Since the global background intensities increase with time (Fig.~\ref{fig:spec_total}), the epoch at which these haloes emerge determines how effectively X-ray feedback operates in different environments.

Fig.~\ref{fig:npop3_low} shows the Pop~III star counts in individual underdense volumes. Although X-ray feedback generally enhances Pop~III star formation in these regions, its effectiveness varies. The observed cases are summarised below.
\begin{enumerate}
    \item \textbf{Volumes~3 and 4.} The X-ray enhancement exceeds the average (factors of 7 and 5, respectively). In the X-ray runs (red lines), Pop~III stars form between $z \sim 16$ and $12$\footnote{An exception occurs in Volume~3, where a Pop~III star forms near the end of the simulation ($z \approx 9$) in a very massive halo ($M \sim 5 \times 10^7~\msun$), likely via atomic rather than \hm cooling. For simplicity, we do not distinguish cooling channels.}, when the X-ray background is near its peak ($\intj_{21} \sim 10^{-5}$). In contrast, only one Pop~III star forms without X-rays (blue line). These regions are dominated by low-mass haloes, making LW suppression particularly efficient and X-ray feedback especially important.
    
    \item \textbf{Volumes~5, 6, and 9.} One or two Pop~III stars form in the X-ray run while none form without it. These regions have relatively few haloes, delaying the emergence of potential Pop~III-forming haloes until $z \sim 12$, at which point weak X-ray feedback is sufficient to trigger star formation in some haloes.

    \item \textbf{Volume~7.} Two Pop~III stars form in the X-ray run versus one in the LW-only case. The first Pop~III star forms at $z \sim 13$, leaving limited time for additional formation, so X-ray feedback is marginal.
   
    \item \textbf{Volume~8.} No Pop~III stars form in either simulation. This region is the most underdense, and candidate haloes emerge only after the LW background becomes strong ($z \lesssim 12$), fully suppressing Pop~III star formation even with X-rays.

    \item \textbf{Volume~10.} One Pop~III star forms in both simulations. The first forms at relatively high redshift ($z \sim 17$), with no further formation at lower redshifts — inconsistent with case (i). We find that the haloes drift towards a massive structure outside the zoom-in region, suggesting numerical artefacts.
\end{enumerate}
These results support our hypothesis that the epoch at which potential Pop~III-forming haloes ($M \sim 10^5$--$10^6~\msun$) first appear governs the efficiency of X-ray feedback. The X-ray intensity peaks near $z \sim 16$ and remains roughly constant until $z \sim 12$, after which Pop~II star formation sharply boosts the LW background. X-ray feedback is therefore most effective when candidate haloes virialise within this redshift window; haloes appearing later are suppressed by the strong LW background.

\begin{figure*}
    \centering
	\includegraphics[width=0.95\textwidth]{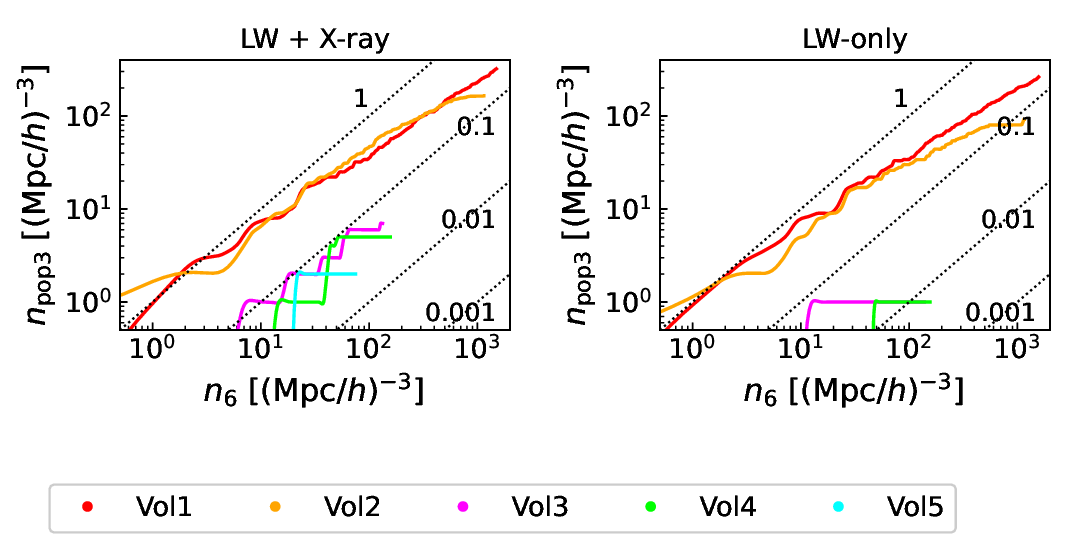}
    \caption{Number density of Pop~III stars ($\npop$) versus that of all haloes more massive than $10^6~\msun$ ($n_{6}$). Results are shown for Volumes 1 to 5 (distinguished by colour) in the X-ray (left panel) and LW-only (right panel) simulations. Dotted lines demarcate ratios 1, 0.1, 0.01, and 0.001. Note that the ratio can exceed unity (see red and orange lines), as Pop~III stars may form in haloes with $M < 10^6~\msun$.}
    \label{fig:nnz}
\end{figure*}

Pop~III star formation in Volumes~1 and 2 can be understood in the same context. In these regions, Pop~III stars begin forming at high redshifts before the X-ray background has fully developed, so $\npop$ in the X-ray vs LW-only runs differ little.

In Section~\ref{sec:mhalo}, we showed that X-rays increase both the number density and occupation fraction of Pop~III-hosting haloes in Volume~2. We now extend this analysis to other simulated volumes. Fig.~\ref{fig:nnz} shows $\npop$ as a function of the number density of haloes more massive than $10^6~\msun$ ($n_6$), a proxy for the number of available Pop~III-forming sites. Since $n_6$ increases monotonically with time, the $x$-axis also traces cosmic evolution.

In Volumes~1 and 2, $\npop/n_6 \approx 1$ for $n_6 < 10$~Mpc$^{-3}$ and approaches $\approx 1/10$ at larger $n_6$, with the X-ray and LW-only runs tracking each other closely. In low-density volumes, however, the ratio in the X-ray runs (left panel) stays near $\npop/n_6 \sim 0.1$, while the LW-only values are ten times lower (Volumes~3 and 4) or zero (Volume~5). This confirms that positive X-ray feedback is more effective in underdense environments.

These results point to a clear prediction of our models: a moderate X-ray background increases $\npop$ by a factor of $\sim 2$ in overdense and mean-density regions, but in underdense regions Pop~III star formation occurs in significant numbers only when X-ray irradiation is present. X-ray feedback is most effective when haloes of mass $\sim 10^6~\msun$ virialise between $z \sim 16$ and $12$, the window during which the X-ray background peaks and the LW background remains moderate. Including additional X-ray sources may shift this window while preserving the overall trend. We leave such investigations for future work.


\section{Discussion}
\label{sec:disc}

\subsection{Comparison to Previous Published Results}
\label{sec:comparison}

Throughout this paper, we have noted that the X-ray-induced increase in $\npop$ ($\eta_{X}$) is substantially smaller than predicted by the semi-analytic model of R16. R16 estimated Pop~III number densities of $\sim 10~(\mpch)^{-3}$ and $\sim 400~(\mpch)^{-3}$ for weak ($\intj_{21} \sim 10^{-8}$)\footnote{Effectively equivalent to the zero-X-ray case in this work.} and moderate ($\intj_{21} \sim 10^{-4}$) X-ray backgrounds, a factor of $\sim 40$ increase. Our simulations yield only a factor of $\sim 2$ in the mean-density environment (Volume~2): $92~(\mpch)^{-3}$ without X-rays versus $165~(\mpch)^{-3}$ with X-rays. The main source of this discrepancy appears to be a significant underestimation of the Pop~III number density for the case without X-rays in R16, by roughly a factor of 10, possibly due to stronger negative LW feedback. The residual discrepancy in the X-ray case (factor of 2--3) likely reflects different physical prescriptions (X-ray and LW mean free paths, \hm cooling rates, spectral hardness) but is most plausibly explained by the lower X-ray intensity reached in our simulations. Our maximum intensity is $\intj_{21} \sim 10^{-5}$ (Fig.~\ref{fig:spec_total}), roughly an order of magnitude below the moderate case in R16. PRS21 showed that dramatic reductions in host halo mass (by $\sim$an order of magnitude) occur only for $\intj_{21} \gtrsim 10^{-4}$, while weaker backgrounds reduce masses by only a few. Since Pop~III supernovae produce a relatively weak X-ray background in our model, the resulting enhancement is moderate and consistent with PRS21.

In Appendix~\ref{sec:xraytest} we examine whether stronger X-ray backgrounds can further increase positive feedback and reproduce the R16 results. Test simulations with artificially enhanced intensities show that $\npop$ increases only moderately with X-ray intensity, because IGM heating introduces a compensating negative feedback — consistent with R16. Increasing the X-ray background by a factor of ten raises $\npop$ by roughly a factor of 2, while a further factor-of-ten increase ($100\times$ fiducial) yields no additional gain. Nevertheless, a factor-of-two increase brings $\npop \sim 300$--$400~(\mpch)^{-3}$, in agreement with R16.

\begin{figure*}
    \centering
	\includegraphics[width=0.95\textwidth]{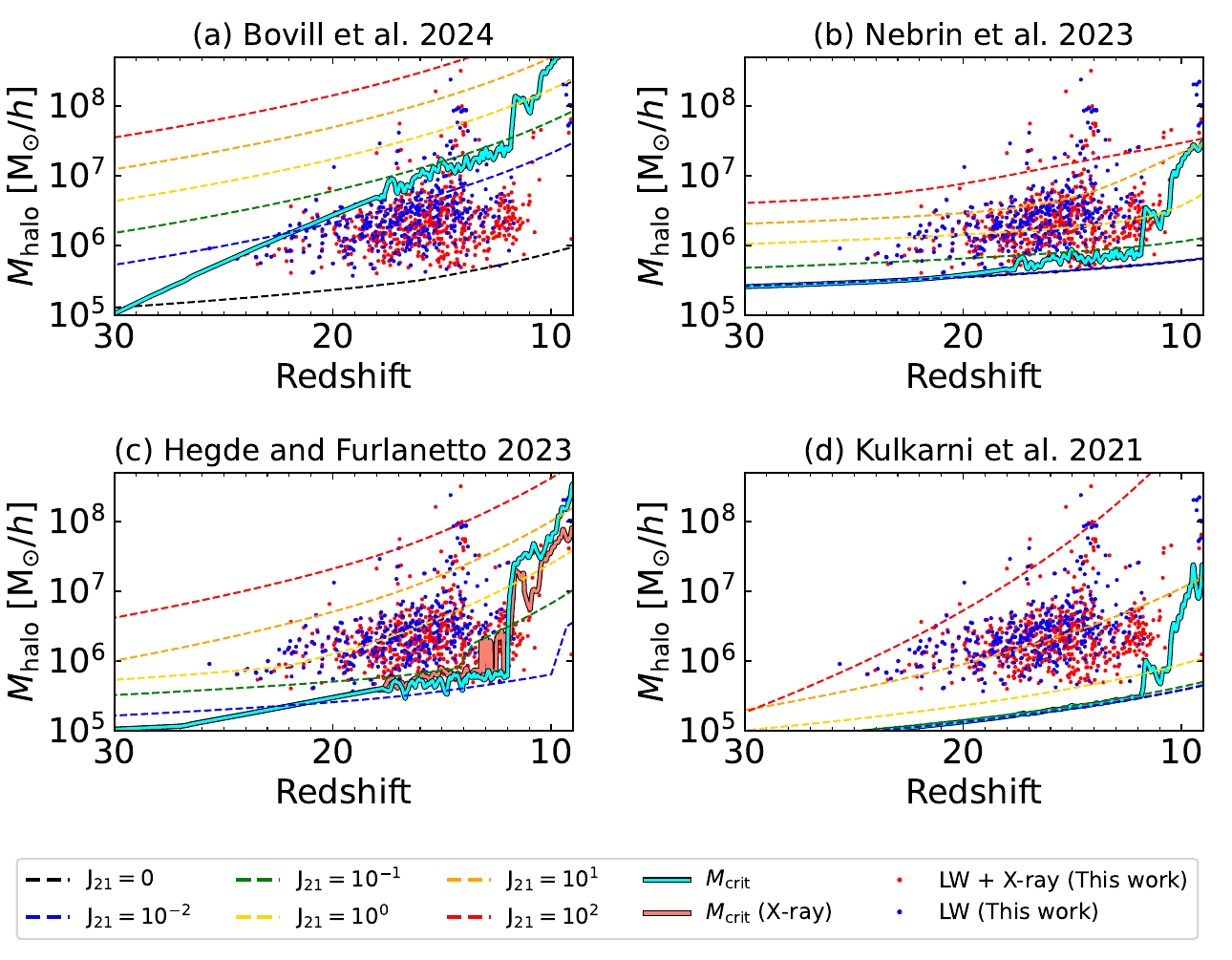}
    \caption{Host halo masses of all simulations compared with the critical masses derived from semi-analytic models \citep{bovill_kindling_2024, nebrin_starbursts_2023, hegde_self-consistent_2023} and numerical simulations \citet{kulkarni_critical_2021} (from Panel~a to d). The dashed lines indicate the critical masses for LW intensities of $\intj_{21} = 0, 10^{-2}, 10^{-1}, 10^{0}, 10^{1}$ and $10^{2}$. The thick cyan line indicates the critical mass evaluated using the average LW intensity in our simulation. \textbf{Panel~a:} Most haloes lie below the blue line ($\intj_{21} = 10^{-2}$), suggesting that LW feedback is stronger in \citet{bovill_kindling_2024} than in our simulations. \textbf{Panel~b:} A non-negligible fraction (76 out of 998) of Pop~III stars lie above the red line ($\intj_{21} = 10^2$), indicating Pop~III star formation in strong local UV backgrounds. Low mass haloes, however, follow the lower bound (cyan line) well. \textbf{Panel~c:} Our low mass haloes also agree well with the minimum masses derived by \citet{hegde_self-consistent_2023}. The cyan and pink lines show the minimum masses derived with their methodology. \textbf{Panel~d:} Halo masses are typically a few times larger than the critical mass (cyan line), suggesting the critical mass is somewhat underestimated in \citet{kulkarni_critical_2021} compared to our simulations.}
    \label{fig:mcrit_sam}
\end{figure*}

\subsubsection{Halo Critical Mass and Fraction of Pop~III-forming Haloes}
\label{sec:mcrit_formula}

The critical halo mass is a key parameter in semi-analytic models of early galaxy formation. In Fig.~\ref{fig:mcrit_sam}, we compare the Pop~III-forming halo masses in our simulations with the critical masses derived in previous studies that examined LW feedback \citep{kulkarni_critical_2021, nebrin_starbursts_2023, bovill_kindling_2024}, and LW + X-ray feedback \citep{hegde_self-consistent_2023} systematically. The thick cyan line shows the critical mass as a function of redshift using the same LW intensity $J_{21}$ as in our fiducial run; dashed lines show critical masses for fixed values of $J_{21}$. In Panel~c, the thick pink line represents the critical mass regulated by LW radiation and X-ray feedback. Overall, our results follow the expected redshift scaling $M_{crit} \propto (1+z)^{-p}$ where $p > 0$ \citep{kulkarni_critical_2021,nebrin_starbursts_2023,bovill_kindling_2024}, but the level of agreement depends on the adopted model.

\begin{enumerate}
    \item \textbf{\citet{bovill_kindling_2024}:} Most of our haloes lie below their predicted threshold, indicating that our Pop~III-forming haloes are less sensitive to LW negative feedback than their model predicts.

    \item \textbf{\citet{nebrin_starbursts_2023}:} Our results are broadly consistent with their model. Most haloes lie above their critical mass, and Pop~III stars form in massive haloes ($M \gtrsim 10^7~\msun$) below $z \approx 12$, when the LW background from Pop~II stars rapidly intensifies. A non-negligible fraction of haloes (76 out of 988, or 7.7\%) lies above the red line ($\intj_{21} = 10^2$), indicating Pop~III star formation in strong local LW backgrounds.

    \item \textbf{\citet{hegde_self-consistent_2023}:} Our haloes masses are broadly consistent with their results, particularly with the minimum masses shown by cyan and pink lines. The critical mass in this case depends on the LW intensity and the electron fraction in the IGM (at mean cosmic density). The critical mass in the X-ray case (pink) remains slightly higher than the LW-only case (cyan) at $z \gtrsim 12$, suggesting positive X-ray feedback is not as effective as in our simulations. At lower redshifts, however, the X-ray regulated critical mass remains lower than the non-X-ray case, showing effective X-ray positive feedback.

    \item \textbf{\citet{kulkarni_critical_2021}:} Our haloes are typically a few times more massive than their predicted critical mass (cyan line), suggesting that LW feedback is more effective at suppressing Pop~III star formation in our model.
\end{enumerate}

By comparing the red and blue points in panel (c) of Fig.~\ref{fig:mcrit_sam} it is clear that at $z<12$ X-rays lower the critical mass in our simulations (see also Fig.~\ref{fig:mhz}). This is qualitatively in agreement with the analytic result shown by the thick pink line in which the X-rays decrease the critical mass. However, the two results show discrepancy at higher redshifts in which the critical mass in the X-ray background remains slightly higher. We verified that enhanced Pop~III star formation in the X-ray background results in an increase in the LW intensity and the subsequent increase in the critical mass (their equation~(29)). Although the enhanced electron fraction lowers the critical mass (see their equations~(43) and (44)), it cannot compensate for the LW feedback as effectively as in our simulations.

The result of \citet{kulkarni_critical_2021} is based on numerical simulations, and we therefore briefly discuss possible origins of the discrepancy between their work and ours. We first note the potential role of DM mass resolution, which is crucial for resolving gas collapse in minihaloes. Our high-resolution runs have a DM particle mass ($\sim 800~\msun$) comparable to that in \citet{kulkarni_critical_2021}, yet our Pop~III-forming haloes are systematically more massive. We therefore conclude that DM mass resolution alone is unlikely to account for the discrepancy.

\begin{figure}
    \centering
	\includegraphics[width=0.48\textwidth]{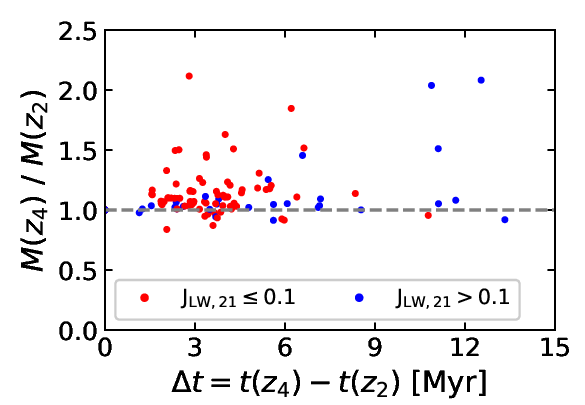}
    \caption{Ratio of halo masses at two epochs ($z_2$ and $z_4$) in our previous simulation (PRS21)). $z_2$ and $z_4$ denote the redshifts at which the central density reaches $10^2$ and $10^4$~\hcc, respectively, while $t_2$ and $t_4$ are the corresponding cosmic ages. Colours indicate the intensity of the LW background. A mass ratio greater than unity indicates that the host halo grows while the central gas collapses and its density increases. Ratios slighly below unity arises from uncertainties in virial mass measurement when halo growth is marginal.}
    \label{fig:delay}
\end{figure}

Another possible origin of the discrepancy is the different Pop~III  formation criteria adopted in the two studies. In this work, we define the critical halo mass at the epoch when a Pop~III star forms from a gas clump with a peak density of $n_{H} > 10^4$~cm$^{-3}$. In contrast, \citet{kulkarni_critical_2021} adopted a density threshold two orders of magnitude lower. We find that this seemingly small difference in definition may contribute to the discrepancy between the two works. To test this, we make use of our previous simulations (PRS21)\footnote{We use these simulations because we recorded the host halo properties when the central gas density increased by an order of magnitude, up to $10^{11}$~\hcc.}. Fig.~\ref{fig:delay} shows the ratio of halo masses at two epochs $z_2$ and $z_4$: corresponding to the redshifts when the central gas density reaches $n_{H} = 10^2$ and $10^4$~\hcc, respectively. These threshold correspond to the Pop~III formation criteria adopted by \citet{kulkarni_critical_2021} and this work, respectively. The figure demonstrates that some haloes may grow in mass by more than a factor of 1.5 (corresponding to $\sim 0.17$~dex) while the central gas density increases by two orders of magnitude from $100$~\hcc. Since LW background delays gas collapse (blue dots), they may enhance the likelihood of substantial halo growth during this phase. This partially explains why haloes in our simulations tend to be higher than the critical mass reported by \citet{kulkarni_critical_2021}. We speculate that this effects, combined with other factors such as numerical diffusion\footnote{In grid-based codes, haloes in larger simulation boxes (e.g., $1$~Mpc$/h$~as adopted in this work) tend to have larger peculiar velocities than those in zoom-in simulations targeting individual haloes. Since numerical diffusion scales with the gas velocity \citep{robertson_computational_2010}, gas in our simulations may remain more diffusive.} and local UV feedback, may delay Pop~III star formation and thereby increase the halo masses. We conclude this section by highlighting that more quantitative and systematic comparisons are required to better understand the discrepancies between simulations.

\subsection{Impacts on the Formation of the Second-generation Stars}
\label{sec:xraypop2}

Pop~III stars supply the first metals needed to initiate Pop~II star formation, thereby regulating the assembly of the first galaxies. Here we discuss how X-ray radiation may indirectly influence early galaxy formation by enhancing metal enrichment in low-density regions that would otherwise remain devoid of stars and metals.

\begin{figure*}
    \centering
	\includegraphics[width=0.95\textwidth]{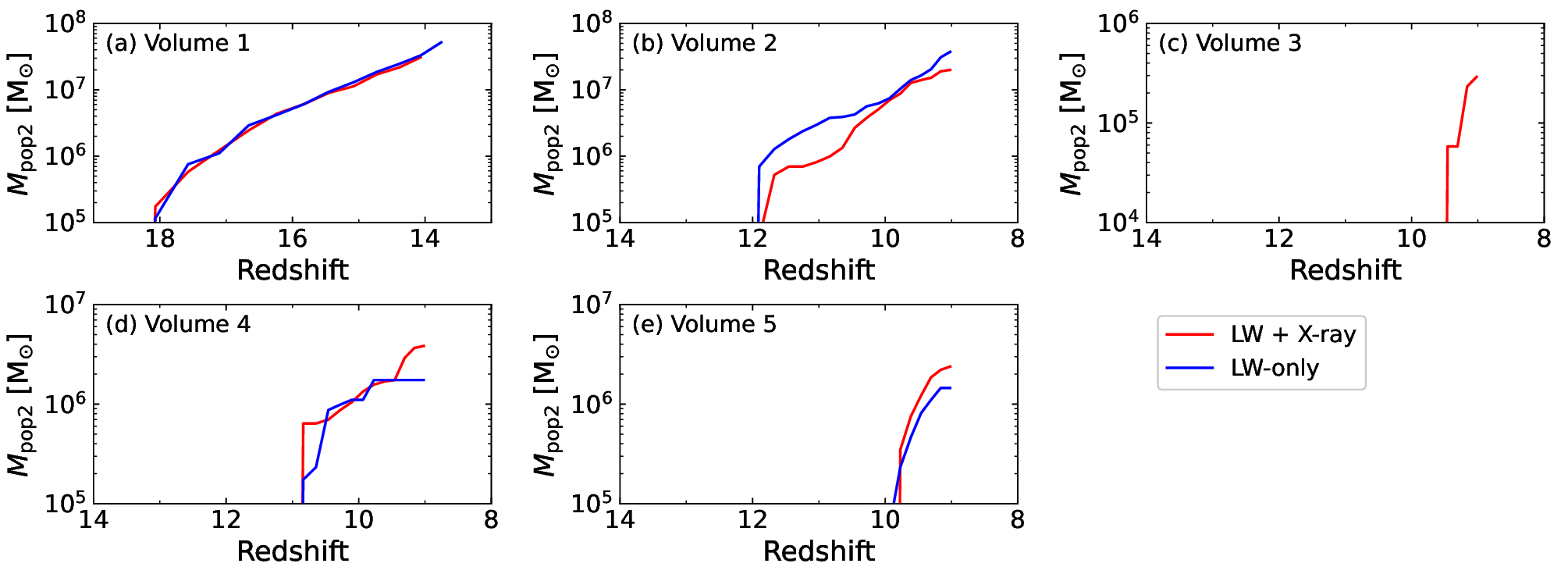}
    \caption{Cumulative mass in Pop~II stars as a function of redshift for Volumes~1 to 5. We show the results of the X-ray runs in red and those of LW-only runs in blue. In Volumes 1 and 2, the mass in Pop~II stars remain larger in the LW-only runs due to stronger Pop~III feedback. In Volumes~3 to 5, on the other hand, the mass in Pop~II stars remain larger in the X-ray runs thanks to the enhanced metal enrichment. In Volume~3, Pop~II stars do not form in the absence of X-rays (i.e., the blue line is not shown).}
    \label{fig:SFRD}
\end{figure*}

Fig.~\ref{fig:SFRD} shows the cumulative Pop~II stellar masses in Volumes~1--5. In Volumes~1 and 2, Pop~II stellar masses are similar in the X-ray and LW-only runs, though slightly larger in the LW-only simulations (blue line). In Volume~2, Pop~II star formation is briefly suppressed between $z=12$ and $z=10$ due to enhanced Pop~III activity in the X-ray run. In contrast, Volumes~3--5 show enhanced Pop~II star formation under X-ray feedback, driven by increased chemical enrichment. Notably, in Volume~3, Pop~II stars form \emph{only} in the presence of X-rays, illustrating the critical role of local metal enrichment.

\begin{figure*}
    \centering
	\includegraphics[width=0.95\textwidth]{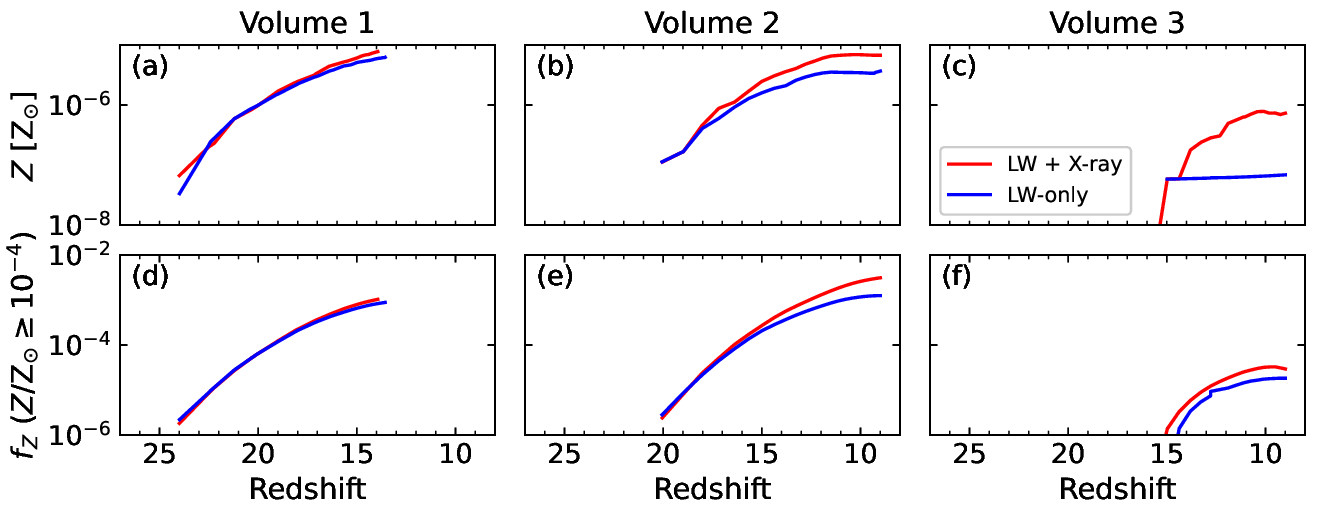}
    \caption{\textbf{Top panels:} Gas metallicity in the target regions of Volume~1, 2, and 3 (Panels~a to c, respectively). A marked difference attributed to the effects of the X-ray background is clear in Volume~3 (Panel~c). \textbf{Bottom panels:} Volume-filling fractions of gas with $Z/$Z$_{\odot} \geq 10^{-4}$. In all three volumes, the X-ray background increases the volume-filling fraction by less than a factor of two. Unlike the pronounced metallicity difference seen in Volume~3 (Panel~c), the change in the volume-filling fraction is modest. The adopted metallicity threshold of $Z = 10^{-4}~\mathrm{Z}_{\odot}$ is somewhat arbitrary, but other thresholds within [$10^{-6}~\mathrm{Z}_{\odot}, 10^{-2}~\mathrm{Z}_{\odot}$] yield qualitatively similar results.}
    \label{fig:volumeZ}
\end{figure*}

Fig.~\ref{fig:volumeZ} examines the impact of X-rays on metal enrichment further, showing the redshift evolution of gas metallicity (top panels) and the volume-filling fraction of gas with $Z > 10^{-4}~\mathrm{Z}_{\odot}$ (bottom panels) for Volumes~1, 2, and 3. In all cases, the X-ray runs (red) show higher metallicities than the LW-only runs (blue), consistent with increased Pop~III activity and subsequent enrichment; this trend holds also for Volumes~4 and 5. The metallicity enhancement is most pronounced in Volume~3, where X-ray feedback is strongest (Panel~c). The bottom panels show that metal-enriched regions are more spatially extended in the X-ray case, though the increase in volume-filling fraction for Volume~3 is smaller than the order-of-magnitude metallicity enhancement in Panel~c. This decoupling — the total metal mass scales with $\npop$, but the enriched volume does not — reflects the local environment: although Volume~3 has a below-average halo number density, Pop~III stars still form in relatively dense locations, so filaments and nearby haloes confine the contaminated gas. By contrast, the single Pop~III star in the LW-only run forms in a more isolated location, allowing metals to spread into voids and occupy a larger fraction of the volume, reducing the contrast in volume-filling fraction between the two runs.

These findings suggest that X-ray backgrounds can significantly shape the stellar masses and metallicities of dwarf galaxies forming in underdense regions. While this work focuses on global properties of the target regions, galaxy-scale characteristics ultimately depend on mass assembly histories and feedback processes. Future work will explore galaxy formation under X-ray feedback using models that include bursty star formation \citep{sugimura_violent_2024, kang_impact_2025} and mechanical supernova feedback \citep{kimm_towards_2015}.

\subsection{Fitting Functions for the Time Evolution of the Pop~III Halo Occupation Probability}
\label{sec:fitting}

\begin{figure*}
    \centering
	\includegraphics[width=0.95\textwidth]{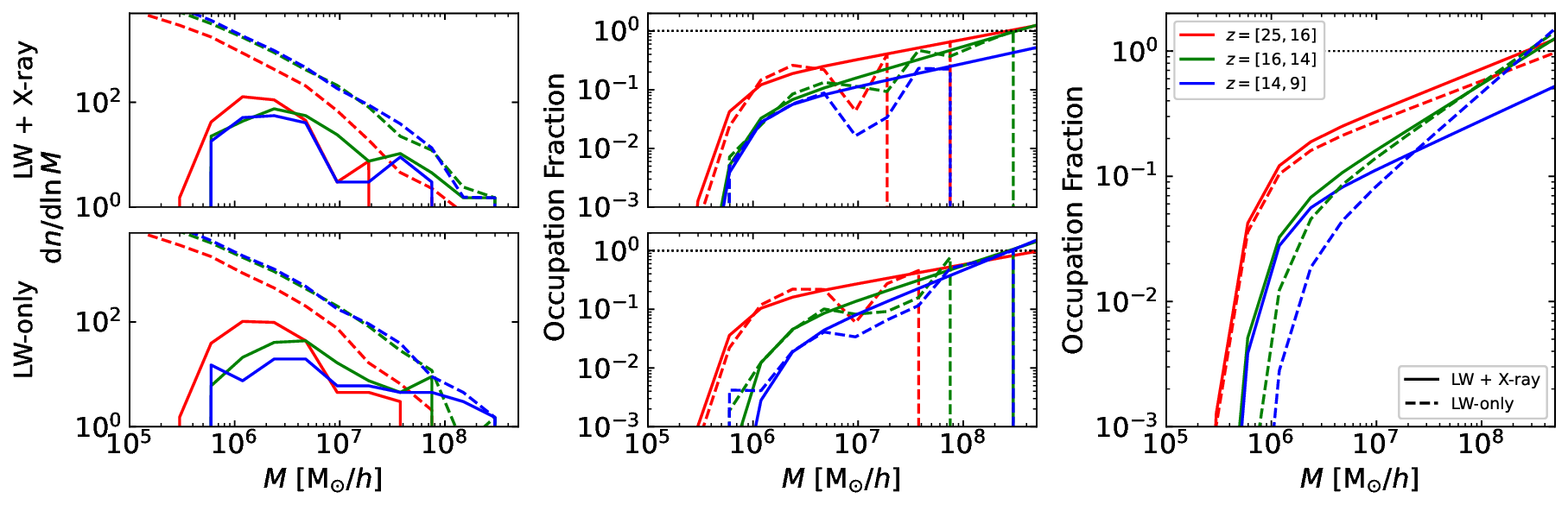}
    \caption{\textbf{Left:} Mass functions of Pop~III-hosting haloes (solid lines) and all haloes (dashed lines) for three redshift ranges: $[25, 16]$, $[16, 14]$, and $[14, 9]$, coloured red, green, and blue, respectively. Mass functions are constructed from the haloes of all 10 volumes. \textbf{Middle:} Occupation fractions (dashed lines) and their fits (solid lines); see the equation in Section~\ref{sec:fitting} for the fitting formula. \textbf{Right:} All fits shown together. X-ray simulation results are shown as solid lines; LW-only results as dashed lines.}
    \label{fig:fitting}
\end{figure*}

We present fitting functions for the Pop~III halo occupation probability using all 10 simulations, irrespective of environment. We construct mass functions of all DM haloes and of Pop~III-hosting haloes for three redshift ranges (see Table~\ref{tab:fitting}). The occupation fraction ($f_{occ}$) in each range is fitted with a power law with an exponential cutoff,
\begin{equation}
    f_{occ} = \min \left\{ 1, ~\left( \frac{M}{M_0} \right)^\beta \exp{ \left[ -\left( \frac{M_{\mathrm{cut}}}{M} \right)^2 \right]}\right\}.
    \label{eq:fitting}
\end{equation}
Here $M_{0}$ is the characteristic halo mass, $\beta$ is the power-law slope, and $M_{\mathrm{cut}}$ is the exponential cutoff mass. Both the power-law and exponential components converge to unity as $M \to M_0$. The best-fit parameter values are given in Table~\ref{tab:fitting} for the three redshift ranges and for the X-ray and LW-only runs. These parameters reflect positive X-ray feedback: lower $\beta$ and $M_{cut}$ reflect increased formation in low-mass haloes (i.e., a reduced critical mass). The difference between the X-ray and LW-only cases is negligible at the highest redshift bin ($z \geq 16$), when the X-ray background is still weak. We provide the halo mass functions, occupation fractions, and fitting functions in Fig.~\ref{fig:fitting}.

\begin{figure*}
    \centering
	\includegraphics[width=0.95\textwidth]{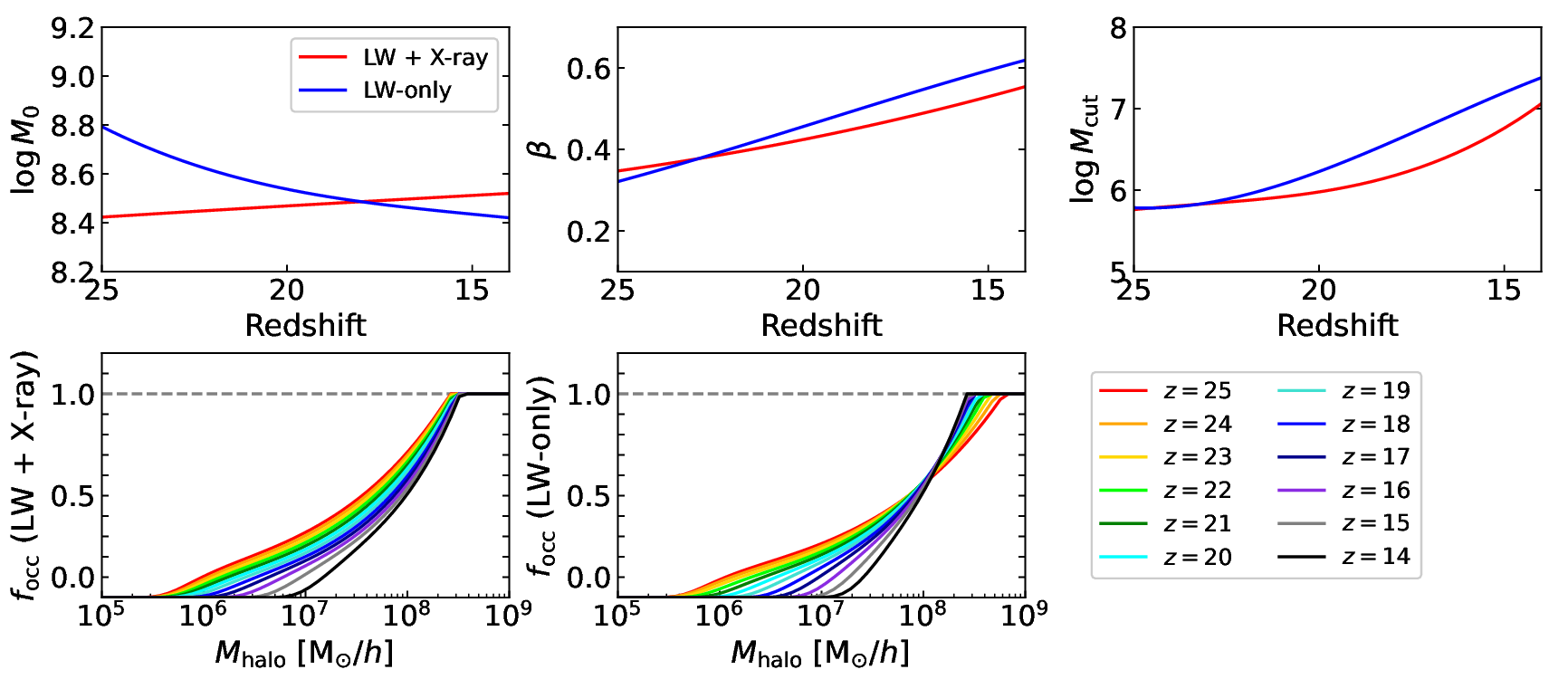}
    \caption{\textbf{Top:} Fits to $\log{M_{0}}, \beta$, and $\log{M_{cut}}$ (equation~(\ref{eq:fitting})).  \textbf{Bottom:} $f_{occ}$ between $z=22$ and $14$ during which Pop~III stars form actively for X-ray (left) and LW-only (right) cases.}
    \label{fig:fitting2}
\end{figure*}

Finally, we also provide formulae for the three parameters as a function of redshift so that $f_{occ}$ can be computed at a given redshift using equation~(\ref{eq:fitting}). $\log{M_{0}}, \beta$, and $\log{M_{cut}}$ are fitted with a cubic polynomial,
\begin{equation}
    \begin{split}
        \log{M_{0}}, \beta, & \log{M_{cut}} = \\
        & p_0 + p_1(z-14) + p_2 (z-14)^2 + p_3 (z-14)^3.
    \end{split}
\end{equation}
Note that this formula is valid for $25 \geq z \geq 14$. We ruled out the lower redshift results as Volume~1 is not considered when constructing the fitting function. The values for the coefficients are given in Table~\ref{tab:fitting2} and are shown in Fig.~\ref{fig:fitting2} as a function of redshift with the occupation fraction estimated.

\begin{table}
    \caption{Parameters for occupation fractions.}
    \centering
    \footnotesize
    \begin{threeparttable}
        \begin{tabular}{ | l | c | c | c |}
		    \hline
            
            LW + X-ray & & & \\
            \hline
            
             & $M_{0}$ & $\beta$ & $M_{cut}$  \\
            \hline
            
            $25 \geq z \geq 16$ & $2.64 \times 10^{8}$ & $0.340$ & $6.24 \times 10^5$ \\
            \hline
            
            $16 \geq z \geq 14$ & $3.21 \times 10^{8}$ & $0.523$ & $8.40 \times 10^5$ \\
            \hline
            
            $14 \geq z \geq 9$ & $2.62 \times 10^{9}$ & $0.391$ & $9.00 \times 10^5$ \\
            \hline

            & & & \\
            \hline
            
            LW-only & & & \\
            \hline
                
            & $M_{0}$ & $\beta$ & $M_{cut}$  \\
            \hline
            
            $25 \geq z \geq 16$ & $5.47 \times 10^{8}$ & $0.323$ & $6.30 \times 10^5$  \\
            \hline
            
            $16 \geq z \geq 14$ & $2.73 \times 10^{8}$ & $0.588$ & $1.31 \times 10^6$  \\
            \hline
            
            $14 \geq z \geq 9$ & $2.85 \times 10^{8}$ & $0.733$ & $1.62 \times 10^6$  \\
            \hline

	\end{tabular}
    \end{threeparttable}
    \label{tab:fitting}
\end{table}

\begin{table}
    \caption{Coefficients of cubic polynomial fit.}
    \centering
    \footnotesize
    \begin{threeparttable}
        \begin{tabular}{ | l | c |}
		    \hline
            
            $\log{M_{0}}$ & $(p_0, p_1, p_2, p_3)$ \\
		    \hline

            LW + X-ray & $(8.52, -8.47 \times 10^{-3}, 5.93 \times 10^{-6}, -3.57 \times 10^{-6})$ \\
            \hline
            
            LW-only & $(8.42, 1.52 \times 10^{-2}, -4.50 \times 10^{-4}, 1.96 \times 10^{-4})$ \\
            \hline

             & \\
            \hline
            
            $\beta$ & \\
		    \hline

            LW + X-ray & $(0.554, -2.54 \times 10^{-2}, 6.50 \times 10^{-4}, -4.68 \times 10^{-6})$ \\
            \hline
            
            LW-only & $(0.619, -2.45 \times 10^{-2}, -7.02 \times 10^{-4}, 4.24 \times 10^{-5})$ \\
            \hline

             & \\
            \hline

            $\log{M_{cut}}$ & \\
		    \hline

            LW + X-ray & $(7.06, -0.331, 3.20 \times 10^{-2}, -1.15 \times 10^{-3})$ \\
            \hline
            
            LW-only & $(7.38, -0.175, -9.50 \times 10^{-3}, 1.11 \times 10^{-3})$ \\

            \hline

	\end{tabular}
    \end{threeparttable}
    \label{tab:fitting2}
\end{table}

\subsection{Impacts of Other X-ray Sources}

Motivated by our previous works (R16; PRS21), the main goal of this work is to investigate the role of Pop~III SNe in the build-up of the global X-ray radiation background and its impact on the the number density of Pop~III haloes. Although we leave the contribution of other X-ray sources to future work, here we speculate on how they may modify our results when included. Several previous studies \citep[e.g.,][]{jeon_radiative_2014,hummel_first_2015} assumed that HMXBs are the primary X-ray sources in the early Universe. Although an individual HMXB emits more X-ray energy in total than a SN\footnote{Assuming that a HMXB accretes at the Eddington rate ($\sim 10^{38}$~erg~s$^{-1}$ per $\msun$) for $\sim 2$~Myr, has a mass of $30~\msun$, and radiates 30~\% of the released energy in X-rays as in previous studies \citep{jeon_radiative_2014,hummel_first_2015}, the total X-ray energy emitted by a HMXB is $6 \times 10^{52}$~erg and is two orders of magnitude greater than X-ray energy emitted by a PISN ($6 \times 10^{50}$~erg, see PR26a).}, they are expected to be less numerous than SNe. For instance, the aforementioned works assumed one HMXB forms per 30 Pop~III binary systems. We therefore speculate that including Pop~III HMXBs would not increase the X-ray background intensity by more than an order of magnitude if we keep the same assumptions. Nevertheless, self-consistent modelling including HMXBs is still warranted because hard X-ray photons may affect gas heating and ionisation differently from soft X-rays emitted by Pop~III SNe.

Although individual Pop~II SNe are thought to be less energetic than Pop~III SNe, Pop~II stars dominate the stellar mass budget by $z \sim 9$ (e.g., $M_{pop3} \sim 10^4~\msun$ versus $M_{pop2} \gtrsim 10^7~\msun$ in Volume~2, see Figs.~\ref{fig:npop3} and \ref{fig:SFRD}). As the dominant mode of star formation transitions from Pop~III to Pop~II at $z \sim 10 - 15$, Pop~II SNe and Pop~II HMXBs may contribute significantly to the build-up of the X-ray background at lower redshifts ($6 \lesssim z \lesssim 9$). This additional X-ray radiation may prolong Pop~III star formation down to the epoch of reionisation ($z \sim 6$) as suggested by \citet{hegde_self-consistent_2023}. Furthermore, this may enable Pop~III star formation in underdense regions where the gas remain pristine, and thereby enhancing the detectability of Pop~III stars or their explosion at lower redshifts \citep{hegde_self-consistent_2023}. Such effects may also influence the formation of dwarf galaxies as suggested in Section~\ref{sec:xraypop2}.

\subsection{Comparison to Empirical Scaling between SFR and X-ray Emission}
\label{sec:obs}

Some semi-analytic models of Pop~III star formation \citep{hegde_self-consistent_2023} adopts a simple scaling relationship between the X-ray luminosity ($L_{X}$) and the SFR observed in local galaxies \citep{lehmer_evolution_2016}, and therefore it is instructive to compare our simulations to these observations.

Although our simulations include X-ray radiative transfer within the computational domain, directly estimating the X-ray luminosities of individual galaxies is not feasible because Pop~III SNe are the only X-ray sources in the simulation. In addition, the low cadence of our simulation  outputs further complicates the estimation of instantaneous X-ray luminosities. For these reasons, we estimate the X-ray luminosities indirectly using simple analytic calculations outlined in Appendix~\ref{sec:Lx}. Here, we focus on the resulting luminosities and their implications, and refer readers to the Appendix for details of the calculations.

\begin{figure*}
    \centering
	\includegraphics[width=0.95\textwidth]{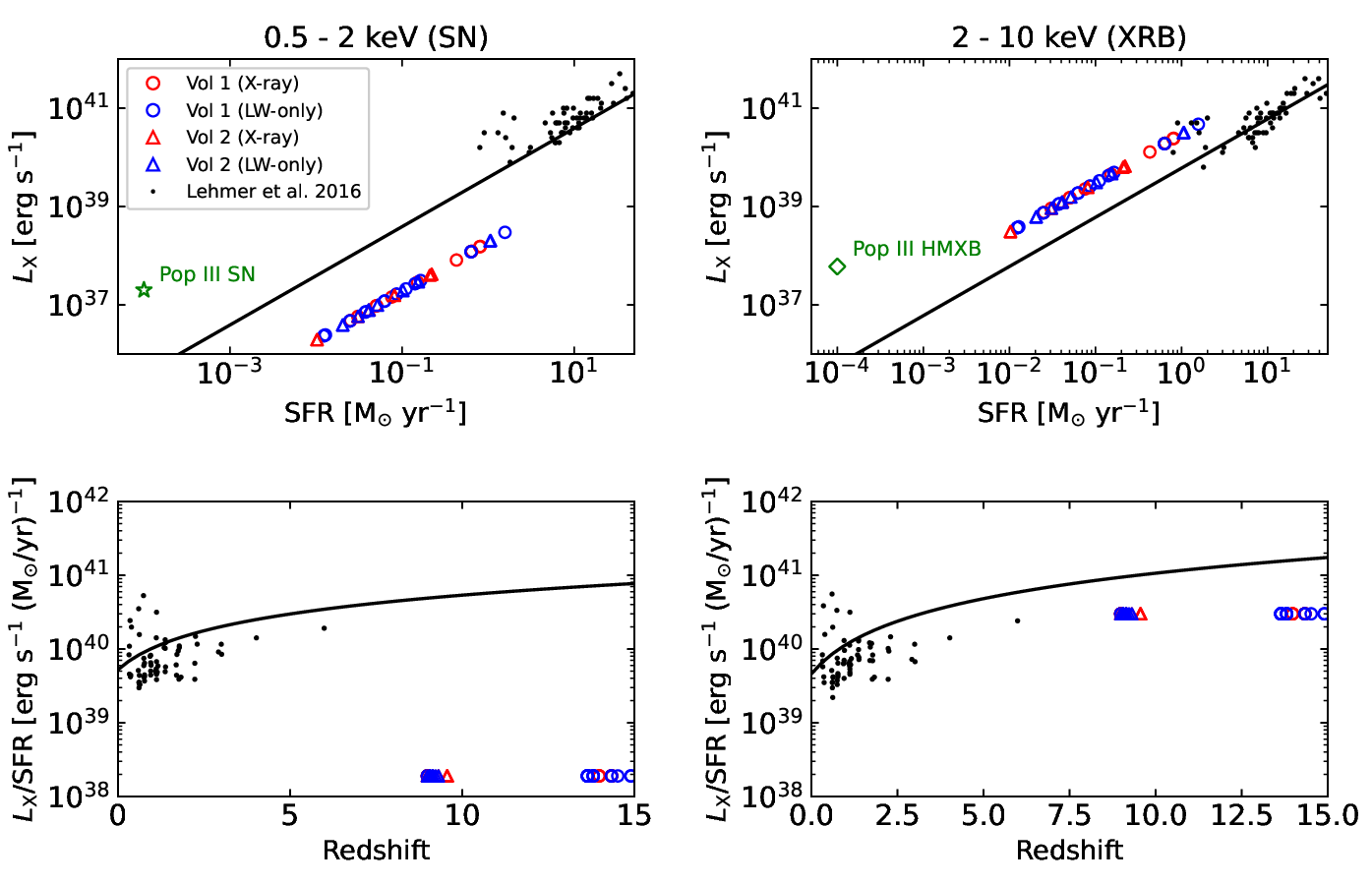}
    \caption{\textbf{Top: } $L_{X}$--SFR relation. In the left panels, we compare X-ray luminosities from Pop~II SNe in our simulated galaxies (circles and triangles) with the observed soft X-ray luminosities of local galaxies from \citep{lehmer_evolution_2016}. Green symbols represent average SFR and X-ray luminosities of X-ray sources associated with Pop~III stars. Black dots represent individual observed galaxies, and the solid line shows their fit (equation~(12) of \citet{lehmer_evolution_2016}). In the right panels, we present the X-ray luminosities obtained when Pop~II X-ray XRBs are assumed to the primary X-ray sources. These results are compared with the observed hard X-ray luminosities of local galaxies. \textbf{Bottom: } $L_{X}/$SFR as a function of redshift. The format is the same as the top panels, but the fits are calculated using equation~(14) of \citet{lehmer_evolution_2016}. We adopt the fits for galaxies with SFR $= 0.3 - 5~\msun$~yr$^{-1}$ (see Fig.~9 of their work) and confirm that all of our galaxies has SFRs in this range.}
    \label{fig:lehmer16}
\end{figure*}

The top panels of Fig.~\ref{fig:lehmer16} compare our estimated galaxy X-ray luminosities (circles and triangles) to the local observational data (black dots) and their fit (black line) derived by \citet{lehmer_evolution_2016}. Our results suggest that at a given SFR the contribution from supernovae is small relative to the observational relationship, whereas the contribution from X-ray binaries (XRBs) is slightly overestimated but most likely accounts for the majority of the X-ray luminosity. In the bottom panels, we present $L_{X}$/SFR as a function of redshift. The bottom left panel shows that the Pop~II SN contribution deviates significantly from the extrapolated fit. Pop~II XRB results shown in the bottom right panel are in closer agreenment with the extrapolated relation, although our predicted luminosities still remain almost an order of magnitude lower. This discrepancy highlights the need for direct high redshift X-ray observations as well as more elaborate models of galaxy formation and X-ray emission. Future work will incorporate more up-to-date physical prescriptions, including metallicity-dependent Pop~II IMFs \citep{chon_transition_2021}, BH accretion models \citep{park_accretion_2011, sugimura_structure_2020}, higher cadence simulation outputs, in order to analyse the X-ray luminosities of galaxies directly from the simulation snapshots.

\subsection{Impacts of Pop~III Initial Mass Function}

In this work, we adopt a fixed Pop~III stellar mass ($M = 100~\msun$) for simplicity and leave the impact of Pop~III initial mass function (IMF) as a future work. Nevertheless, we briefly discuss its possible implications in this section.

Pop~III stars are thought to radiate at the Eddington luminosity $L \sim 10^{38}$ ergs s$^{-1}$ $(M/\msun)$ \citep{bromm_generic_2001}, and the LW luminosity in the model we adopted \citep{schaerer_properties_2002} is also proportional to the mass of the star. This implies that, for a given stellar mass budget, the slope of the IMF (top-heavy versus bottom-heavy) does not significantly affect the intensity of the LW background because it primarily depends on the total stellar mass.
If the hypernovae/PISN energy is assumed to be constant, the IMF plays an important role in the build-up of the X-ray background because it regulates the multiplicity of Pop~III stars and therefore the number of SNe. On the other hand, if the SN energy scales with Pop~III mass as in \citet{heger_nucleosynthetic_2002}, the IMF is not expected to affect our results significantly because both LW and X-ray backgrounds are primarily determined by total stellar mass.

We point out that the IMF affects not only the luminosities of Pop~III stars but also their fates. According to \citet{heger_nucleosynthetic_2002}, a non-rotating Pop~III star may explode as a PISN ($140~\msun \leq M \leq 260~\msun$) or collapse directly into BHs without producing energetic supernovae ($40~\msun \leq M \leq 140~\msun$ or $M \geq 260~\msun$). We emphasize that these mass windows are intrinsically uncertain and apply only to non-rotating stars. However, taken at face value, if the IMF of Pop~III stars has a high mass cutoff at $M<140~\msun$ or if it is sufficiently top-heavy such that most Pop~III stars fall within the direct-collapse mass ranges, the contribution of Pop~III SNe to the X-ray background could be substantially reduced as well as their metal enrichment, which is necessary to transition to Pop~II star formation. Another complication is that the Pop~III IMF itself may depend on the X-ray background. Although we assumed a fixed Pop~III mass, \cite{park_population_2021b} found that an effect of X-rays is to reduce the mass and multiplicity of Pop~III stars. This implies that X-ray feedback loop may weaken over time if the supernova energy and the number of HMXBs (and therefore X-ray energy) decreases as the X-ray background builds up. This effect will be incorporated self-consistently in future work.


\section{Summary}
\label{sec:summary}

An X-ray background in the early Universe partially ionises primordial gas and boosts \hm formation, thereby lowering the threshold for Pop~III star formation in low-mass dark matter minihaloes (R16; PRS21). Here we quantify this effect using cosmological simulations across a range of cosmic environments. Such simulations are challenging, as they require both high resolution to resolve the first stars and a large volume to self-consistently estimate the radiation backgrounds produced by distant sources. In the companion paper (PR26a), we describe our methodology for estimating the X-ray and LW backgrounds produced by Pop~III stars and their PISNe, including the self-consistent treatment of the radiation feedback loop and the on-the-fly calculation of backgrounds from sources outside the computational box. The X-ray background is estimated using Volume~2, a representative volume whose halo mass function closely matches the cosmic mean at $z = 9$, and is then applied as a spatially uniform field to regions with different halo number densities. 

We find that Pop~III supernovae produce a weak ($\intj_{21} \sim 10^{-5}$) soft ($E \sim 0.2$--$2.0$~keV) X-ray background alongside a moderate LW background ($\intj_{21} \sim 10^{-2}$--$10^{-1}$) by $z \approx 12$. Below $z \approx 12$, the onset of Pop~II star formation intensifies the LW background to $\intj_{21} \sim 10^{1}$.

In this paper, we examine the effects of the LW and X-ray backgrounds produced by Pop~III and their PISNe on the requirements for Pop~III star formation and their consequent number density. The key results are as follows.
\begin{enumerate}
    \item {\bf Overall effects of X-rays.} The X-ray background produces mild positive feedback, promoting Pop~III star formation in two ways. First, it reduces the mean host halo mass by up to a factor of two. Second, it enables Pop~III star formation in haloes that would otherwise fail to form stars, increasing the halo occupation probability. When the critical mass is defined as the mass at which the occupation probability reaches 50\% \citep{kulkarni_critical_2021}, the X-ray background reduces this mass by a factor of $2$--$3$ at $z = 9$, increasing $\npop$ by a similar factor.

    \item {\bf Enhanced Pop~III formation in underdense regions.} X-rays boost $\npop$ in all environments, but the magnitude depends on halo number density. The increase is smallest in overdense regions (Volume~1, factor of 1.2) and strongest in underdense regions, reaching $\sim 3$ on average and up to 7 (Volume~3). Without X-rays, Pop~III stars cannot form at all in Volume~3.

    \item {\bf Reasons for the strong environmental dependence.} The dependence is primarily governed by the halo growth history. Pop~III star formation is promoted when candidate haloes ($M \gtrsim 10^6~\msun$) emerge between $z \approx 16$ and $z \approx 12$, when the X-ray background peaks ($\intj_{21} \sim 10^{-5}$) and the LW background remains moderate ($\intj_{21} \sim 10^{-1}$). At redshifts $z \lesssim 12$, the strong LW background suppresses Pop~III star formation. In overdense regions (Volumes~1 and 2), Pop~III stars form earlier, when the X-ray background is still weak, limiting its impact.

    \item {\bf Impact on Pop~III detection at lower redshifts and on the first galaxies.} We observe enhanced Pop~II star formation and higher gas metallicity in the low-density IGM as a consequence of Pop~III formation in underdense regions enabled by X-ray irradiation. Mean-density and overdense regions (Volumes~1 and 2) show little change in Pop~II star formation, while underdense regions (Volumes~3, 4, and 5) exhibit larger Pop~II masses and higher metallicity in the presence of X-rays. This suggests that the clustering of dwarf galaxies at high redshift may be shaped by an early X-ray background. In addition, X-ray effects are strongest in models where Pop~III stars continue forming to relatively low redshift, as late formation occurs in pockets of low-metallicity IGM in underdense environments. We speculate that including X-rays in conjunction with Pop~II star formation may further sustain Pop~III star formation in underdense regions at late times.

    \item {\bf Fitting functions for the Pop~III occupation probability of minihaloes.} We provide fitting functions for the redshift evolution of the Pop~III halo occupation probability for simulations with and without X-ray feedback. These can be used in semi-analytic models of Pop~III star formation and for predictions of PISNe rates, particularly in light of upcoming Roman and LSST large-scale surveys. Given our limited sample size, we are unable to split the fitting functions by environment.
\end{enumerate}

Future work will incorporate additional X-ray sources (HMXBs, AGNs, Pop~II supernovae, etc.) and more sophisticated physical prescriptions — including bursty star formation \citep{sugimura_violent_2024,kang_impact_2025}, mechanical supernova feedback \citep{kimm_towards_2015}, metallicity-dependent Pop~II IMFs \citep{chon_transition_2021}, and BH gas accretion \citep{park_accretion_2011,sugimura_structure_2020} — to obtain more robust predictions for $\npop$ and to explore the role of X-rays in the formation of the first galaxies.

\section*{Acknowledgements}

We acknowledge the anonymous referee for constructive comments and helping us improve the quality of the paper. JP acknowledges the support of the NRF of Korea grant funded by the Korean government (MSIT, RS-2022-NR068800). This work was supported by the BK21 Fostering Outstanding Universities for Research (FOUR) Program (4120200513819). The authors acknowledge the University of Maryland supercomputing resources (\url{http://hpcc.umd.edu}) made available for conducting this research. This work was granted access to the HPC resources of KISTI under allocation KSC-2024-CRE-0200. Large data transfers were supported by KREONET, managed and operated by KISTI.

\section*{Data Availability}

The data underlying this article will be shared on reasonable request to the corresponding author.



\bibliographystyle{mnras}
\bibliography{reference} 

@article{abe_formation_2021,
	title = {Formation of the first galaxies in the aftermath of the first supernovae},
	volume = {508},
	issn = {0035-8711},
	url = {https://ui.adsabs.harvard.edu/abs/2021MNRAS.508.3226A},
	doi = {10.1093/mnras/stab2637},
	urldate = {2025-03-09},
	journal = {\mnras},
	author = {Abe, Makito and Yajima, Hidenobu and Khochfar, Sadegh and Dalla Vecchia, Claudio and Omukai, Kazuyuki},
	month = dec,
	year = {2021},
	note = {Publisher: OUP
ADS Bibcode: 2021MNRAS.508.3226A},
	pages = {3226--3238},
}

@article{abel_formation_2002,
	title = {The {Formation} of the {First} {Star} in the {Universe}},
	volume = {295},
	issn = {0036-8075},
	url = {https://ui.adsabs.harvard.edu/abs/2002Sci...295...93A/abstract},
	doi = {10.1126/science.1063991},
	language = {en},
	number = {5552},
	urldate = {2025-10-04},
	journal = {Science},
	author = {Abel, Tom and Bryan, Greg L. and Norman, Michael L.},
	month = jan,
	year = {2002},
	pages = {93--98},
}

@article{ahn_does_2007,
	title = {Does radiative feedback by the first stars promote or prevent second generation star formation?},
	volume = {375},
	issn = {0035-8711},
	url = {https://ui.adsabs.harvard.edu/abs/2007MNRAS.375..881A/abstract},
	doi = {10.1111/j.1365-2966.2006.11332.x},
	language = {en},
	number = {3},
	urldate = {2025-09-09},
	journal = {\mnras},
	author = {Ahn, Kyungjin and Shapiro, Paul R.},
	month = mar,
	year = {2007},
	pages = {881--908},
}

@article{ahn_spatially_2015,
	title = {Spatially {Extended} 21 cm {Signal} from {Strongly} {Clustered} {Uv} and {X}-{Ray} {Sources} in the {Early} {Universe}},
	volume = {802},
	issn = {0004-637X},
	url = {https://ui.adsabs.harvard.edu/abs/2015ApJ...802....8A},
	doi = {10.1088/0004-637X/802/1/8},
	urldate = {2024-02-16},
	journal = {\apj},
	author = {Ahn, Kyungjin and Xu, Hao and Norman, Michael L. and Alvarez, Marcelo A. and Wise, John H.},
	month = mar,
	year = {2015},
	note = {ADS Bibcode: 2015ApJ...802....8A},
	pages = {8},
}

@article{behroozi_rockstar_2013,
	title = {The {ROCKSTAR} {Phase}-space {Temporal} {Halo} {Finder} and the {Velocity} {Offsets} of {Cluster} {Cores}},
	volume = {762},
	issn = {0004-637X},
	url = {https://ui.adsabs.harvard.edu/abs/2013ApJ...762..109B},
	doi = {10.1088/0004-637X/762/2/109},
	urldate = {2024-02-02},
	journal = {\apj},
	author = {Behroozi, Peter S. and Wechsler, Risa H. and Wu, Hao-Yi},
	month = jan,
	year = {2013},
	note = {ADS Bibcode: 2013ApJ...762..109B},
	pages = {109},
}

@article{bleuler_phew_2015,
	title = {{PHEW}: a parallel segmentation algorithm for three-dimensional {AMR} datasets. {Application} to structure detection in self-gravitating flows},
	volume = {2},
	shorttitle = {{PHEW}},
	url = {https://ui.adsabs.harvard.edu/abs/2015ComAC...2....5B},
	doi = {10.1186/s40668-015-0009-7},
	urldate = {2022-08-23},
	journal = {ComAC},
	author = {Bleuler, Andreas and Teyssier, Romain and Carassou, Sébastien and Martizzi, Davide},
	month = jun,
	year = {2015},
	note = {ADS Bibcode: 2015ComAC...2....5B},
	pages = {5},
}

@article{bouwens_newly_2019,
	title = {Newly {Discovered} {Bright} z âˆ¼ 9-10 {Galaxies} and {Improved} {Constraints} on {Their} {Prevalence} {Using} the {Full} {CANDELS} {Area}},
	volume = {880},
	issn = {0004-637X},
	url = {https://ui.adsabs.harvard.edu/abs/2019ApJ...880...25B},
	doi = {10.3847/1538-4357/ab24c5},
	urldate = {2023-06-06},
	journal = {\apj},
	author = {Bouwens, R. J. and Stefanon, M. and Oesch, P. A. and Illingworth, G. D. and Nanayakkara, T. and Roberts-Borsani, G. and LabbÃ©, I. and Smit, R.},
	month = jul,
	year = {2019},
	note = {ADS Bibcode: 2019ApJ...880...25B},
	pages = {25},
}

@article{bovill_kindling_2024,
	title = {Kindling the {First} {Stars}. {I}. {Dependence} of {Detectability} of the {First} {Stars} with {JWST} on the {Population} {III} {Stellar} {Masses}},
	volume = {962},
	issn = {0004-637X},
	url = {https://ui.adsabs.harvard.edu/abs/2024ApJ...962...49B/abstract},
	doi = {10.3847/1538-4357/ad148a},
	language = {en},
	number = {1},
	urldate = {2025-06-06},
	journal = {\apj},
	author = {Bovill, Mia Sauda and Stiavelli, Massimo and Wiggins, Alessa Ibrahim and Ricotti, Massimo and Trenti, Michele},
	month = feb,
	year = {2024},
	pages = {49},
}

@article{bromm_generic_2001,
	title = {Generic {Spectrum} and {Ionization} {Efficiency} of a {Heavy} {Initial} {Mass} {Function} for the {First} {Stars}},
	volume = {552},
	issn = {0004-637X},
	url = {https://ui.adsabs.harvard.edu/abs/2001ApJ...552..464B},
	doi = {10.1086/320549},
	urldate = {2023-06-27},
	journal = {\apj},
	author = {Bromm, Volker and Kudritzki, Rolf P. and Loeb, Abraham},
	month = may,
	year = {2001},
	note = {ADS Bibcode: 2001ApJ...552..464B},
	pages = {464--472},
}

@article{bromm_formation_2002,
	title = {The {Formation} of the {First} {Stars}. {I}. {The} {Primordial} {Star}-forming {Cloud}},
	volume = {564},
	issn = {0004-637X},
	url = {https://ui.adsabs.harvard.edu/abs/2002ApJ...564...23B},
	doi = {10.1086/323947},
	urldate = {2023-09-16},
	journal = {\apj},
	author = {Bromm, Volker and Coppi, Paolo S. and Larson, Richard B.},
	month = jan,
	year = {2002},
	note = {ADS Bibcode: 2002ApJ...564...23B},
	pages = {23--51},
}

@article{carniani_spectroscopic_2024,
	title = {Spectroscopic confirmation of two luminous galaxies at a redshift of 14},
	volume = {633},
	issn = {0028-0836},
	url = {https://ui.adsabs.harvard.edu/abs/2024Natur.633..318C/abstract},
	doi = {10.1038/s41586-024-07860-9},
	language = {en},
	number = {8029},
	urldate = {2025-10-04},
	journal = {Nature},
	author = {Carniani, Stefano and Hainline, Kevin and D'Eugenio, Francesco and Eisenstein, Daniel J. and Jakobsen, Peter and Witstok, Joris and Johnson, Benjamin D. and Chevallard, Jacopo and Maiolino, Roberto and Helton, Jakob M. and Willott, Chris and Robertson, Brant and Alberts, Stacey and Arribas, Santiago and Baker, William M. and Bhatawdekar, Rachana and Boyett, Kristan and Bunker, Andrew J. and Cameron, Alex J. and Cargile, Phillip A. and Charlot, Stéphane and Curti, Mirko and Curtis-Lake, Emma and Egami, Eiichi and Giardino, Giovanna and Isaak, Kate and Ji, Zhiyuan and Jones, Gareth C. and Kumari, Nimisha and Maseda, Michael V. and Parlanti, Eleonora and Pérez-González, Pablo G. and Rawle, Tim and Rieke, George and Rieke, Marcia and Del Pino, Bruno Rodríguez and Saxena, Aayush and Scholtz, Jan and Smit, Renske and Sun, Fengwu and Tacchella, Sandro and Übler, Hannah and Venturi, Giacomo and Williams, Christina C. and Willmer, Christopher N. A.},
	month = sep,
	year = {2024},
	pages = {318},
}

@article{chiaki_metal-poor_2018,
	title = {Metal-poor star formation triggered by the feedback effects from {Pop} {III} stars},
	volume = {475},
	issn = {0035-8711},
	url = {https://ui.adsabs.harvard.edu/abs/2018MNRAS.475.4378C},
	doi = {10.1093/mnras/sty040},
	urldate = {2022-12-06},
	journal = {\mnras},
	author = {Chiaki, Gen and Susa, Hajime and Hirano, Shingo},
	month = apr,
	year = {2018},
	note = {ADS Bibcode: 2018MNRAS.475.4378C},
	pages = {4378--4395},
}

@article{chon_forming_2019,
	title = {Forming {Pop} {III} binaries in self-gravitating discs: how to keep the orbital angular momentum},
	volume = {488},
	issn = {0035-8711},
	shorttitle = {Forming {Pop} {III} binaries in self-gravitating discs},
	url = {https://ui.adsabs.harvard.edu/abs/2019MNRAS.488.2658C},
	doi = {10.1093/mnras/stz1824},
	urldate = {2022-08-06},
	journal = {\mnras},
	author = {Chon, Sunmyon and Hosokawa, Takashi},
	month = sep,
	year = {2019},
	note = {ADS Bibcode: 2019MNRAS.488.2658C},
	pages = {2658--2672},
}

@article{chon_transition_2021,
	title = {Transition of the initial mass function in the metal-poor environments},
	volume = {508},
	issn = {0035-8711},
	url = {https://ui.adsabs.harvard.edu/abs/2021MNRAS.508.4175C},
	doi = {10.1093/mnras/stab2497},
	urldate = {2022-08-06},
	journal = {\mnras},
	author = {Chon, Sunmyon and Omukai, Kazuyuki and Schneider, Raffaella},
	month = dec,
	year = {2021},
	note = {ADS Bibcode: 2021MNRAS.508.4175C},
	pages = {4175--4192},
}

@article{clark_formation_2011,
	title = {The {Formation} and {Fragmentation} of {Disks} {Around} {Primordial} {Protostars}},
	volume = {331},
	issn = {0036-8075},
	url = {https://ui.adsabs.harvard.edu/abs/2011Sci...331.1040C},
	doi = {10.1126/science.1198027},
	urldate = {2022-12-06},
	journal = {Science},
	author = {Clark, Paul C. and Glover, Simon C. O. and Smith, Rowan J. and Greif, Thomas H. and Klessen, Ralf S. and Bromm, Volker},
	month = feb,
	year = {2011},
	note = {ADS Bibcode: 2011Sci...331.1040C},
	pages = {1040},
}

@article{clark_gravitational_2011,
	title = {Gravitational {Fragmentation} in {Turbulent} {Primordial} {Gas} and the {Initial} {Mass} {Function} of {Population} {III} {Stars}},
	volume = {727},
	issn = {0004-637X},
	url = {https://ui.adsabs.harvard.edu/abs/2011ApJ...727..110C},
	doi = {10.1088/0004-637X/727/2/110},
	urldate = {2023-07-07},
	journal = {\apj},
	author = {Clark, Paul C. and Glover, Simon C. O. and Klessen, Ralf S. and Bromm, Volker},
	month = feb,
	year = {2011},
	note = {ADS Bibcode: 2011ApJ...727..110C},
	pages = {110},
}

@article{curtis-lake_spectroscopic_2023,
	title = {Spectroscopic confirmation of four metal-poor galaxies at z = 10.3-13.2},
	volume = {7},
	issn = {2397-3366},
	url = {https://ui.adsabs.harvard.edu/abs/2023NatAs...7..622C},
	doi = {10.1038/s41550-023-01918-w},
	urldate = {2023-10-20},
	journal = {Nature Astronomy},
	author = {Curtis-Lake, Emma and Carniani, Stefano and Cameron, Alex and Charlot, Stephane and Jakobsen, Peter and Maiolino, Roberto and Bunker, Andrew and Witstok, Joris and Smit, Renske and Chevallard, Jacopo and Willott, Chris and Ferruit, Pierre and Arribas, Santiago and Bonaventura, Nina and Curti, Mirko and D'Eugenio, Francesco and Franx, Marijn and Giardino, Giovanna and Looser, Tobias J. and LÃ¼tzgendorf, Nora and Maseda, Michael V. and Rawle, Tim and Rix, Hans-Walter and RodrÃ­guez del Pino, Bruno and Ãœbler, Hannah and Sirianni, Marco and Dressler, Alan and Egami, Eiichi and Eisenstein, Daniel J. and Endsley, Ryan and Hainline, Kevin and Hausen, Ryan and Johnson, Benjamin D. and Rieke, Marcia and Robertson, Brant and Shivaei, Irene and Stark, Daniel P. and Tacchella, Sandro and Williams, Christina C. and Willmer, Christopher N. A. and Bhatawdekar, Rachana and Bowler, Rebecca and Boyett, Kristan and Chen, Zuyi and de Graaff, Anna and Helton, Jakob M. and Hviding, Raphael E. and Jones, Gareth C. and Kumari, Nimisha and Lyu, Jianwei and Nelson, Erica and Perna, Michele and Sandles, Lester and Saxena, Aayush and Suess, Katherine A. and Sun, Fengwu and Topping, Michael W. and Wallace, Imaan E. B. and Whitler, Lily},
	month = may,
	year = {2023},
	note = {ADS Bibcode: 2023NatAs...7..622C},
	pages = {622--632},
}

@article{diemer_colossus_2018,
	title = {{COLOSSUS}: {A} {Python} {Toolkit} for {Cosmology}, {Large}-scale {Structure}, and {Dark} {Matter} {Halos}},
	volume = {239},
	issn = {0067-0049},
	shorttitle = {{COLOSSUS}},
	url = {https://ui.adsabs.harvard.edu/abs/2018ApJS..239...35D},
	doi = {10.3847/1538-4365/aaee8c},
	urldate = {2024-02-02},
	journal = {\apjs},
	author = {Diemer, Benedikt},
	month = dec,
	year = {2018},
	note = {ADS Bibcode: 2018ApJS..239...35D},
	pages = {35},
}

@article{dubois_agn-driven_2013,
	title = {{AGN}-driven quenching of star formation: morphological and dynamical implications for early-type galaxies},
	volume = {433},
	issn = {0035-8711},
	shorttitle = {{AGN}-driven quenching of star formation},
	url = {https://ui.adsabs.harvard.edu/abs/2013MNRAS.433.3297D/abstract},
	doi = {10.1093/mnras/stt997},
	language = {en},
	number = {4},
	urldate = {2026-05-28},
	journal = {\mnras},
	author = {Dubois, Yohan and Gavazzi, Raphaël and Peirani, Sébastien and Silk, Joseph},
	month = aug,
	year = {2013},
	pages = {3297--3313},
}

@article{finkelstein_census_2022,
	title = {A {Census} of the {Bright} z = 8.5-11 {Universe} with the {Hubble} and {Spitzer} {Space} {Telescopes} in the {CANDELS} {Fields}},
	volume = {928},
	issn = {0004-637X},
	url = {https://ui.adsabs.harvard.edu/abs/2022ApJ...928...52F},
	doi = {10.3847/1538-4357/ac3aed},
	urldate = {2023-06-06},
	journal = {\apj},
	author = {Finkelstein, Steven L. and Bagley, Micaela and Song, Mimi and Larson, Rebecca and Papovich, Casey and Dickinson, Mark and Finkelstein, Keely D. and Koekemoer, Anton M. and Pirzkal, Norbert and Somerville, Rachel S. and Yung, L. Y. Aaron and Behroozi, Peter and Ferguson, Harry and Giavalisco, Mauro and Grogin, Norman and Hathi, Nimish and Hutchison, Taylor A. and Jung, Intae and Kocevski, Dale and Kawinwanichakij, Lalitwadee and Rojas-Ruiz, SofÃ­a and Ryan, Russell and Snyder, Gregory F. and Tacchella, Sandro},
	month = mar,
	year = {2022},
	note = {ADS Bibcode: 2022ApJ...928...52F},
	pages = {52},
}

@article{finkelstein_ceers_2023,
	title = {{CEERS} {Key} {Paper}. {I}. {An} {Early} {Look} into the {First} 500 {Myr} of {Galaxy} {Formation} with {JWST}},
	volume = {946},
	issn = {0004-637X},
	url = {https://ui.adsabs.harvard.edu/abs/2023ApJ...946L..13F},
	doi = {10.3847/2041-8213/acade4},
	urldate = {2023-10-03},
	journal = {\apj},
	author = {Finkelstein, Steven L. and Bagley, Micaela B. and Ferguson, Henry C. and Wilkins, Stephen M. and Kartaltepe, Jeyhan S. and Papovich, Casey and Yung, L. Y. Aaron and Haro, Pablo Arrabal and Behroozi, Peter and Dickinson, Mark and Kocevski, Dale D. and Koekemoer, Anton M. and Larson, Rebecca L. and Le Bail, AurÃ©lien and Morales, Alexa M. and PÃ©rez-GonzÃ¡lez, Pablo G. and Burgarella, Denis and DavÃ©, Romeel and Hirschmann, Michaela and Somerville, Rachel S. and Wuyts, Stijn and Bromm, Volker and Casey, Caitlin M. and Fontana, Adriano and Fujimoto, Seiji and Gardner, Jonathan P. and Giavalisco, Mauro and Grazian, Andrea and Grogin, Norman A. and Hathi, Nimish P. and Hutchison, Taylor A. and Jha, Saurabh W. and Jogee, Shardha and Kewley, Lisa J. and Kirkpatrick, Allison and Long, Arianna S. and Lotz, Jennifer M. and Pentericci, Laura and Pierel, Justin D. R. and Pirzkal, Nor and Ravindranath, Swara and Ryan, Russell E. and Trump, Jonathan R. and Yang, Guang and Bhatawdekar, Rachana and Bisigello, Laura and Buat, VÃ©ronique and CalabrÃ², Antonello and Castellano, Marco and Cleri, Nikko J. and Cooper, M. C. and Croton, Darren and Daddi, Emanuele and Dekel, Avishai and Elbaz, David and Franco, Maximilien and Gawiser, Eric and Holwerda, Benne W. and Huertas-Company, Marc and Jaskot, Anne E. and Leung, Gene C. K. and Lucas, Ray A. and Mobasher, Bahram and Pandya, Viraj and Tacchella, Sandro and Weiner, Benjamin J. and Zavala, Jorge A.},
	month = mar,
	year = {2023},
	note = {ADS Bibcode: 2023ApJ...946L..13F},
	pages = {L13},
}

@article{glover_radiative_2003,
	title = {Radiative feedback from an early {X}-ray background},
	volume = {340},
	issn = {0035-8711},
	url = {https://ui.adsabs.harvard.edu/abs/2003MNRAS.340..210G},
	doi = {10.1046/j.1365-8711.2003.06311.x},
	urldate = {2023-02-08},
	journal = {\mnras},
	author = {Glover, S. C. O. and Brand, P. W. J. L.},
	month = mar,
	year = {2003},
	note = {ADS Bibcode: 2003MNRAS.340..210G},
	pages = {210--226},
}

@article{greif_simulations_2011,
	title = {Simulations on a {Moving} {Mesh}: {The} {Clustered} {Formation} of {Population} {III} {Protostars}},
	volume = {737},
	issn = {0004-637X},
	shorttitle = {Simulations on a {Moving} {Mesh}},
	url = {https://ui.adsabs.harvard.edu/abs/2011ApJ...737...75G},
	doi = {10.1088/0004-637X/737/2/75},
	urldate = {2022-12-06},
	journal = {\apj},
	author = {Greif, Thomas H. and Springel, Volker and White, Simon D. M. and Glover, Simon C. O. and Clark, Paul C. and Smith, Rowan J. and Klessen, Ralf S. and Bromm, Volker},
	month = aug,
	year = {2011},
	note = {ADS Bibcode: 2011ApJ...737...75G},
	pages = {75},
}

@article{greif_formation_2012,
	title = {Formation and evolution of primordial protostellar systems},
	volume = {424},
	issn = {0035-8711},
	url = {https://ui.adsabs.harvard.edu/abs/2012MNRAS.424..399G},
	doi = {10.1111/j.1365-2966.2012.21212.x},
	urldate = {2023-04-18},
	journal = {\mnras},
	author = {Greif, Thomas H. and Bromm, Volker and Clark, Paul C. and Glover, Simon C. O. and Smith, Rowan J. and Klessen, Ralf S. and Yoshida, Naoki and Springel, Volker},
	month = jul,
	year = {2012},
	note = {ADS Bibcode: 2012MNRAS.424..399G},
	pages = {399--415},
}

@article{hahn_multi-scale_2011,
	title = {Multi-scale initial conditions for cosmological simulations},
	volume = {415},
	issn = {0035-8711},
	url = {https://ui.adsabs.harvard.edu/abs/2011MNRAS.415.2101H},
	doi = {10.1111/j.1365-2966.2011.18820.x},
	urldate = {2023-06-09},
	journal = {\mnras},
	author = {Hahn, Oliver and Abel, Tom},
	month = aug,
	year = {2011},
	note = {ADS Bibcode: 2011MNRAS.415.2101H},
	pages = {2101--2121},
}

@article{haiman_radiative_2000,
	title = {The {Radiative} {Feedback} of the {First} {Cosmological} {Objects}},
	volume = {534},
	issn = {0004-637X},
	url = {https://ui.adsabs.harvard.edu/abs/2000ApJ...534...11H},
	doi = {10.1086/308723},
	urldate = {2022-07-06},
	journal = {\apj},
	author = {Haiman, ZoltÃ¡n and Abel, Tom and Rees, Martin J.},
	month = may,
	year = {2000},
	note = {ADS Bibcode: 2000ApJ...534...11H},
	pages = {11--24},
}

@article{haardt_radiative_2012,
	title = {Radiative {Transfer} in a {Clumpy} {Universe}. {IV}. {New} {Synthesis} {Models} of the {Cosmic} {UV}/{X}-{Ray} {Background}},
	volume = {746},
	issn = {0004-637X},
	url = {https://ui.adsabs.harvard.edu/abs/2012ApJ...746..125H},
	doi = {10.1088/0004-637X/746/2/125},
	urldate = {2023-06-26},
	journal = {\apj},
	author = {Haardt, Francesco and Madau, Piero},
	month = feb,
	year = {2012},
	note = {ADS Bibcode: 2012ApJ...746..125H},
	pages = {125},
}

@article{hegde_self-consistent_2023,
	title = {A self-consistent semi-analytic model for {Population} {III} star formation in minihaloes},
	volume = {525},
	issn = {0035-8711},
	url = {https://ui.adsabs.harvard.edu/abs/2023MNRAS.525..428H/abstract},
	doi = {10.1093/mnras/stad2308},
	language = {en},
	number = {1},
	urldate = {2026-05-28},
	journal = {\mnras},
	author = {Hegde, Sahil and Furlanetto, Steven R.},
	month = oct,
	year = {2023},
	pages = {428--447},
}

@article{hegde_efficient_2025,
	title = {Efficient semi-analytic modelling of {Pop} {III} star formation from {Cosmic} {Dawn} to {Reionization}},
	volume = {8},
	issn = {2565-6120},
	url = {https://ui.adsabs.harvard.edu/abs/2025OJAp....8E.147H/abstract},
	doi = {10.33232/001c.145070},
	language = {en},
	urldate = {2026-05-28},
	journal = {Open J. Astrophys.},
	author = {Hegde, Sahil and Furlanetto, Steven R.},
	month = oct,
	year = {2025},
	pages = {147},
}

@article{heger_nucleosynthetic_2002,
	title = {The {Nucleosynthetic} {Signature} of {Population} {III}},
	volume = {567},
	issn = {0004-637X},
	url = {https://ui.adsabs.harvard.edu/abs/2002ApJ...567..532H},
	doi = {10.1086/338487},
	urldate = {2022-07-06},
	journal = {\apj},
	author = {Heger, A. and Woosley, S. E.},
	month = mar,
	year = {2002},
	note = {ADS Bibcode: 2002ApJ...567..532H},
	pages = {532--543},
}

@article{hirano_one_2014,
	title = {One {Hundred} {First} {Stars}: {Protostellar} {Evolution} and the {Final} {Masses}},
	volume = {781},
	issn = {0004-637X},
	shorttitle = {One {Hundred} {First} {Stars}},
	url = {https://ui.adsabs.harvard.edu/abs/2014ApJ...781...60H},
	doi = {10.1088/0004-637X/781/2/60},
	urldate = {2023-02-13},
	journal = {\apj},
	author = {Hirano, Shingo and Hosokawa, Takashi and Yoshida, Naoki and Umeda, Hideyuki and Omukai, Kazuyuki and Chiaki, Gen and Yorke, Harold W.},
	month = feb,
	year = {2014},
	note = {ADS Bibcode: 2014ApJ...781...60H},
	pages = {60},
}

@article{hirano_primordial_2015,
	title = {Primordial star formation under the influence of far ultraviolet radiation: 1540 cosmological haloes and the stellar mass distribution},
	volume = {448},
	issn = {0035-8711},
	shorttitle = {Primordial star formation under the influence of far ultraviolet radiation},
	url = {https://ui.adsabs.harvard.edu/abs/2015MNRAS.448..568H},
	doi = {10.1093/mnras/stv044},
	urldate = {2022-07-06},
	journal = {\mnras},
	author = {Hirano, S. and Hosokawa, T. and Yoshida, N. and Omukai, K. and Yorke, H. W.},
	month = mar,
	year = {2015},
	note = {ADS Bibcode: 2015MNRAS.448..568H},
	pages = {568--587},
}

@article{hirano_formation_2017,
	title = {Formation and survival of {Population} {III} stellar systems},
	volume = {470},
	issn = {0035-8711},
	url = {https://ui.adsabs.harvard.edu/abs/2017MNRAS.470..898H},
	doi = {10.1093/mnras/stx1220},
	urldate = {2022-07-08},
	journal = {\mnras},
	author = {Hirano, Shingo and Bromm, Volker},
	month = sep,
	year = {2017},
	note = {ADS Bibcode: 2017MNRAS.470..898H},
	pages = {898--914},
}

@article{hummel_first_2015,
	title = {The first stars: formation under {X}-ray feedback},
	volume = {453},
	issn = {0035-8711},
	shorttitle = {The first stars},
	url = {https://ui.adsabs.harvard.edu/abs/2015MNRAS.453.4136H},
	doi = {10.1093/mnras/stv1902},
	urldate = {2022-07-06},
	journal = {\mnras},
	author = {Hummel, Jacob A. and Stacy, Athena and Jeon, Myoungwon and Oliveri, Anthony and Bromm, Volker},
	month = nov,
	year = {2015},
	note = {ADS Bibcode: 2015MNRAS.453.4136H},
	pages = {4136--4147},
}

@article{incatasciato_modelling_2023,
	title = {Modelling the cosmological {Lyman}-{Werner} background radiation field in the early {Universe}},
	volume = {522},
	issn = {0035-8711},
	url = {https://ui.adsabs.harvard.edu/abs/2023MNRAS.522..330I},
	doi = {10.1093/mnras/stad1008},
	urldate = {2025-05-02},
	journal = {\mnras},
	author = {Incatasciato, Andrea and Khochfar, Sadegh and Oñorbe, Jose},
	month = jun,
	year = {2023},
	note = {Publisher: OUP
ADS Bibcode: 2023MNRAS.522..330I},
	pages = {330--349},
}

@article{jeon_radiative_2014,
	title = {Radiative feedback from high-mass {X}-ray binaries on the formation of the first galaxies and early reionization},
	volume = {440},
	issn = {0035-8711},
	url = {https://ui.adsabs.harvard.edu/abs/2014MNRAS.440.3778J},
	doi = {10.1093/mnras/stu444},
	urldate = {2022-07-06},
	journal = {\mnras},
	author = {Jeon, Myoungwon and Pawlik, Andreas H. and Bromm, Volker and Milosavljević, Miloš},
	month = jun,
	year = {2014},
	note = {ADS Bibcode: 2014MNRAS.440.3778J},
	pages = {3778--3796},
}

@article{jeon_recovery_2014,
	title = {Recovery from {Population} {III} supernova explosions and the onset of second-generation star formation},
	volume = {444},
	issn = {0035-8711},
	url = {https://ui.adsabs.harvard.edu/abs/2014MNRAS.444.3288J},
	doi = {10.1093/mnras/stu1980},
	urldate = {2022-07-06},
	journal = {\mnras},
	author = {Jeon, Myoungwon and Pawlik, Andreas H. and Bromm, Volker and Milosavljević, Miloš},
	month = nov,
	year = {2014},
	note = {ADS Bibcode: 2014MNRAS.444.3288J},
	pages = {3288--3300},
}

@article{kang_impact_2025,
	title = {Impact of star formation models on the growth of simulated galaxies at high redshifts},
	volume = {693},
	issn = {0004-6361},
	url = {https://ui.adsabs.harvard.edu/abs/2025A&A...693A.149K/abstract},
	doi = {10.1051/0004-6361/202451502},
	language = {en},
	urldate = {2025-11-04},
	journal = {\aap},
	author = {Kang, Cheonsu and Kimm, Taysun and Han, Daniel and Katz, Harley and Devriendt, Julien and Slyz, Adrianne and Teyssier, Romain},
	month = jan,
	year = {2025},
	pages = {A149},
}

@article{kimm_towards_2015,
	title = {Towards simulating star formation in turbulent high-z galaxies with mechanical supernova feedback},
	volume = {451},
	issn = {0035-8711},
	url = {https://ui.adsabs.harvard.edu/abs/2015MNRAS.451.2900K/abstract},
	doi = {10.1093/mnras/stv1211},
	language = {en},
	number = {3},
	urldate = {2025-11-04},
	journal = {\mnras},
	author = {Kimm, Taysun and Cen, Renyue and Devriendt, Julien and Dubois, Yohan and Slyz, Adrianne},
	month = aug,
	year = {2015},
	pages = {2900--2921},
}

@article{kroupa_variation_2001,
	title = {On the variation of the initial mass function},
	volume = {322},
	issn = {0035-8711},
	url = {https://ui.adsabs.harvard.edu/abs/2001MNRAS.322..231K/abstract},
	doi = {10.1046/j.1365-8711.2001.04022.x},
	language = {en},
	number = {2},
	urldate = {2026-05-28},
	journal = {\mnras},
	author = {Kroupa, Pavel},
	month = apr,
	year = {2001},
	pages = {231--246},
}

@article{kulkarni_critical_2021,
	title = {The {Critical} {Dark} {Matter} {Halo} {Mass} for {Population} {III} {Star} {Formation}: {Dependence} on {Lyman}-{Werner} {Radiation}, {Baryon}-dark {Matter} {Streaming} {Velocity}, and {Redshift}},
	volume = {917},
	issn = {0004-637X},
	shorttitle = {The {Critical} {Dark} {Matter} {Halo} {Mass} for {Population} {III} {Star} {Formation}},
	url = {https://ui.adsabs.harvard.edu/abs/2021ApJ...917...40K/abstract},
	doi = {10.3847/1538-4357/ac08a3},
	language = {en},
	number = {1},
	urldate = {2025-05-06},
	journal = {\apj},
	author = {Kulkarni, Mihir and Visbal, Eli and Bryan, Greg L.},
	month = aug,
	year = {2021},
	pages = {40},
}

@article{lehmer_evolution_2016,
	title = {The {Evolution} of {Normal} {Galaxy} {X}-{Ray} {Emission} through {Cosmic} {History}: {Constraints} from the 6 {MS} {Chandra} {Deep} {Field}-{South}},
	volume = {825},
	issn = {0004-637X},
	shorttitle = {The {Evolution} of {Normal} {Galaxy} {X}-{Ray} {Emission} through {Cosmic} {History}},
	url = {https://ui.adsabs.harvard.edu/abs/2016ApJ...825....7L/abstract},
	doi = {10.3847/0004-637X/825/1/7},
	language = {en},
	number = {1},
	urldate = {2026-05-24},
	journal = {\apj},
	author = {Lehmer, B. D. and Basu-Zych, A. R. and Mineo, S. and Brandt, W. N. and Eufrasio, R. T. and Fragos, T. and Hornschemeier, A. E. and Luo, B. and Xue, Y. Q. and Bauer, F. E. and Gilfanov, M. and Ranalli, P. and Schneider, D. P. and Shemmer, O. and Tozzi, P. and Trump, J. R. and Vignali, C. and Wang, J.-X. and Yukita, M. and Zezas, A.},
	month = jul,
	year = {2016},
	pages = {7},
}

@article{machacek_simulations_2001,
	title = {Simulations of {Pregalactic} {Structure} {Formation} with {Radiative} {Feedback}},
	volume = {548},
	issn = {0004-637X},
	url = {https://ui.adsabs.harvard.edu/abs/2001ApJ...548..509M/abstract},
	doi = {10.1086/319014},
	language = {en},
	number = {2},
	urldate = {2026-05-29},
	journal = {\apj},
	author = {Machacek, Marie E. and Bryan, Greg L. and Abel, Tom},
	month = feb,
	year = {2001},
	pages = {509--521},
}

@article{machacek_effects_2003,
	title = {Effects of a soft {X}-ray background on structure formation at high redshift},
	volume = {338},
	issn = {0035-8711},
	url = {https://ui.adsabs.harvard.edu/abs/2003MNRAS.338..273M/abstract},
	doi = {10.1046/j.1365-8711.2003.06054.x},
	language = {en},
	number = {2},
	urldate = {2026-05-29},
	journal = {\mnras},
	author = {Machacek, M. E. and Bryan, G. L. and Abel, T.},
	month = jan,
	year = {2003},
	pages = {273--286},
}

@article{madau_radiation_2017,
	title = {Radiation {Backgrounds} at {Cosmic} {Dawn}: {X}-{Rays} from {Compact} {Binaries}},
	volume = {840},
	issn = {0004-637X},
	shorttitle = {Radiation {Backgrounds} at {Cosmic} {Dawn}},
	url = {https://ui.adsabs.harvard.edu/abs/2017ApJ...840...39M/abstract},
	doi = {10.3847/1538-4357/aa6af9},
	language = {en},
	number = {1},
	urldate = {2025-06-25},
	journal = {\apj},
	author = {Madau, Piero and Fragos, Tassos},
	month = may,
	year = {2017},
	pages = {39},
}

@article{nebrin_starbursts_2023,
	title = {Starbursts in low-mass haloes at {Cosmic} {Dawn}. {I}. {The} critical halo mass for star formation},
	volume = {524},
	issn = {0035-8711},
	url = {https://ui.adsabs.harvard.edu/abs/2023MNRAS.524.2290N/abstract},
	doi = {10.1093/mnras/stad1852},
	language = {en},
	number = {2},
	urldate = {2025-06-06},
	journal = {\mnras},
	author = {Nebrin, Olof and Giri, Sambit K. and Mellema, Garrelt},
	month = sep,
	year = {2023},
	pages = {2290},
}

@article{oshea_population_2008,
	title = {Population {III} {Star} {Formation} in a Î›{CDM} {Universe}. {II}. {Effects} of a {Photodissociating} {Background}},
	volume = {673},
	issn = {0004-637X},
	url = {https://ui.adsabs.harvard.edu/abs/2008ApJ...673...14O},
	doi = {10.1086/524006},
	urldate = {2023-06-09},
	journal = {\apj},
	author = {O'Shea, Brian W. and Norman, Michael L.},
	month = jan,
	year = {2008},
	note = {ADS Bibcode: 2008ApJ...673...14O},
	pages = {14--33},
}

@article{park_accretion_2011,
	title = {Accretion onto {Intermediate}-mass {Black} {Holes} {Regulated} by {Radiative} {Feedback}. {I}. {Parametric} {Study} for {Spherically} {Symmetric} {Accretion}},
	volume = {739},
	issn = {0004-637X},
	url = {https://ui.adsabs.harvard.edu/abs/2011ApJ...739....2P},
	doi = {10.1088/0004-637X/739/1/2},
	urldate = {2024-01-04},
	journal = {\apj},
	author = {Park, KwangHo and Ricotti, Massimo},
	month = sep,
	year = {2011},
	note = {ADS Bibcode: 2011ApJ...739....2P},
	pages = {2},
}

@article{park_population_2021a,
	title = {Population {III} star formation in an {X}-ray background - {I}. {Critical} halo mass of formation and total mass in stars},
	volume = {508},
	issn = {0035-8711},
	url = {https://ui.adsabs.harvard.edu/abs/2021MNRAS.508.6176P},
	doi = {10.1093/mnras/stab2999},
	urldate = {2022-07-06},
	journal = {\mnras},
	author = {Park, Jongwon and Ricotti, Massimo and Sugimura, Kazuyuki},
	month = dec,
	year = {2021},
	note = {ADS Bibcode: 2021MNRAS.508.6176P},
	pages = {6176--6192},
}

@article{park_population_2021b,
	title = {Population {III} star formation in an {X}-ray background - {II}. {Protostellar} discs, multiplicity, and mass function of the stars},
	volume = {508},
	issn = {0035-8711},
	url = {https://ui.adsabs.harvard.edu/abs/2021MNRAS.508.6193P},
	doi = {10.1093/mnras/stab3000},
	urldate = {2022-07-06},
	journal = {\mnras},
	author = {Park, Jongwon and Ricotti, Massimo and Sugimura, Kazuyuki},
	month = dec,
	year = {2021},
	note = {ADS Bibcode: 2021MNRAS.508.6193P},
	pages = {6193--6208},
}

@article{park_population_2023,
	title = {Population {III} star formation in an {X}-ray background: {III}. {Periodic} radiative feedback and luminosity induced by elliptical orbits},
	volume = {521},
	issn = {0035-8711},
	shorttitle = {Population {III} star formation in an {X}-ray background},
	url = {https://ui.adsabs.harvard.edu/abs/2023MNRAS.521.5334P},
	doi = {10.1093/mnras/stad895},
	urldate = {2023-04-18},
	journal = {\mnras},
	author = {Park, Jongwon and Ricotti, Massimo and Sugimura, Kazuyuki},
	month = jun,
	year = {2023},
	note = {ADS Bibcode: 2023MNRAS.521.5334P},
	pages = {5334--5353},
}

@article{park_origin_2024,
	title = {On the origin of outward migration of {Population} {III} stars},
	volume = {528},
	issn = {0035-8711},
	url = {https://ui.adsabs.harvard.edu/abs/2024MNRAS.528.6895P},
	doi = {10.1093/mnras/stae518},
	urldate = {2024-02-27},
	journal = {\mnras},
	author = {Park, Jongwon and Ricotti, Massimo and Sugimura, Kazuyuki},
	month = mar,
	year = {2024},
	note = {ADS Bibcode: 2024MNRAS.528.6895P},
	pages = {6895--6914},
}

@article{park_population_2026,
	title = {Population {III} star formation in an {X}-ray background: {IV}. {On}-the-fly calculation of radiation backgrounds and their impact on the intergalactic medium},
	shorttitle = {Population {III} star formation in an {X}-ray background},
	url = {https://ui.adsabs.harvard.edu/abs/2026arXiv260326352P/abstract},
	doi = {10.48550/arXiv.2603.26352},
	language = {en},
	urldate = {2026-06-04},
	journal = {arXiv e-prints},
	author = {Park, Jongwon and Ricotti, Massimo},
	month = mar,
	year = {2026},
	pages = {arXiv:2603.26352},
}

@article{power_primordial_2009,
	title = {Primordial globular clusters, {X}-ray binaries and cosmological reionization},
	volume = {395},
	issn = {0035-8711},
	url = {https://ui.adsabs.harvard.edu/abs/2009MNRAS.395.1146P/abstract},
	doi = {10.1111/j.1365-2966.2009.14628.x},
	language = {en},
	number = {2},
	urldate = {2026-05-28},
	journal = {\mnras},
	author = {Power, C. and Wynn, G. A. and Combet, C. and Wilkinson, M. I.},
	month = may,
	year = {2009},
	pages = {1146--1152},
}

@article{regan_emergence_2020,
	title = {The emergence of the first star-free atomic cooling haloes in the {Universe}},
	volume = {492},
	issn = {0035-8711},
	url = {https://ui.adsabs.harvard.edu/abs/2020MNRAS.492.3021R},
	doi = {10.1093/mnras/staa035},
	urldate = {2023-06-09},
	journal = {\mnras},
	author = {Regan, John A. and Wise, John H. and O'Shea, Brian W. and Norman, Michael L.},
	month = feb,
	year = {2020},
	note = {ADS Bibcode: 2020MNRAS.492.3021R},
	pages = {3021--3031},
}

@article{ricotti_fate_2002a,
	title = {The {Fate} of the {First} {Galaxies}. {I}. {Self}-consistent {Cosmological} {Simulations} with {Radiative} {Transfer}},
	volume = {575},
	issn = {0004-637X},
	url = {https://ui.adsabs.harvard.edu/abs/2002ApJ...575...33R},
	doi = {10.1086/341255},
	urldate = {2022-12-06},
	journal = {\apj},
	author = {Ricotti, Massimo and Gnedin, Nickolay Y. and Shull, J. Michael},
	month = aug,
	year = {2002},
	note = {ADS Bibcode: 2002ApJ...575...33R},
	pages = {33--48},
}

@article{ricotti_fate_2002b,
	title = {The {Fate} of the {First} {Galaxies}. {II}. {Effects} of {Radiative} {Feedback}},
	volume = {575},
	issn = {0004-637X},
	url = {https://ui.adsabs.harvard.edu/abs/2002ApJ...575...49R},
	doi = {10.1086/341256},
	urldate = {2022-12-06},
	journal = {\apj},
	author = {Ricotti, Massimo and Gnedin, Nickolay Y. and Shull, J. Michael},
	month = aug,
	year = {2002},
	note = {ADS Bibcode: 2002ApJ...575...49R},
	pages = {49--67},
}

@article{ricotti_x-ray_2016,
	title = {X-ray twinkles and {Population} {III} stars},
	volume = {462},
	issn = {0035-8711},
	url = {https://ui.adsabs.harvard.edu/abs/2016MNRAS.462..601R},
	doi = {10.1093/mnras/stw1672},
	urldate = {2022-07-06},
	journal = {\mnras},
	author = {Ricotti, Massimo},
	month = oct,
	year = {2016},
	note = {ADS Bibcode: 2016MNRAS.462..601R},
	pages = {601--609},
}

@article{robertson_computational_2010,
	title = {Computational {Eulerian} hydrodynamics and {Galilean} invariance},
	volume = {401},
	issn = {0035-8711},
	url = {https://ui.adsabs.harvard.edu/abs/2010MNRAS.401.2463R/abstract},
	doi = {10.1111/j.1365-2966.2009.15823.x},
	language = {en},
	number = {4},
	urldate = {2026-05-27},
	journal = {\mnras},
	author = {Robertson, Brant E. and Kravtsov, Andrey V. and Gnedin, Nickolay Y. and Abel, Tom and Rudd, Douglas H.},
	month = feb,
	year = {2010},
	pages = {2463--2476},
}

@article{robertson_earliest_2024,
	title = {Earliest {Galaxies} in the {JADES} {Origins} {Field}: {Luminosity} {Function} and {Cosmic} {Star} {Formation} {Rate} {Density} 300 {Myr} after the {Big} {Bang}},
	volume = {970},
	issn = {0004-637X},
	shorttitle = {Earliest {Galaxies} in the {JADES} {Origins} {Field}},
	url = {https://ui.adsabs.harvard.edu/abs/2024ApJ...970...31R/abstract},
	doi = {10.3847/1538-4357/ad463d},
	language = {en},
	number = {1},
	urldate = {2025-10-04},
	journal = {\apj},
	author = {Robertson, Brant and Johnson, Benjamin D. and Tacchella, Sandro and Eisenstein, Daniel J. and Hainline, Kevin and Arribas, Santiago and Baker, William M. and Bunker, Andrew J. and Carniani, Stefano and Cargile, Phillip A. and Carreira, Courtney and Charlot, Stephane and Chevallard, Jacopo and Curti, Mirko and Curtis-Lake, Emma and D'Eugenio, Francesco and Egami, Eiichi and Hausen, Ryan and Helton, Jakob M. and Jakobsen, Peter and Ji, Zhiyuan and Jones, Gareth C. and Maiolino, Roberto and Maseda, Michael V. and Nelson, Erica and Pérez-González, Pablo G. and Puskás, Dávid and Rieke, Marcia and Smit, Renske and Sun, Fengwu and Übler, Hannah and Whitler, Lily and Williams, Christina C. and Willmer, Christopher N. A. and Willott, Chris and Witstok, Joris},
	month = jul,
	year = {2024},
	pages = {31},
}

@article{rosdahl_ramses-rt_2013,
	title = {{RAMSES}-{RT}: radiation hydrodynamics in the cosmological context},
	volume = {436},
	issn = {0035-8711},
	shorttitle = {{RAMSES}-{RT}},
	url = {https://ui.adsabs.harvard.edu/abs/2013MNRAS.436.2188R},
	doi = {10.1093/mnras/stt1722},
	urldate = {2022-07-08},
	journal = {\mnras},
	author = {Rosdahl, J. and Blaizot, J. and Aubert, D. and Stranex, T. and Teyssier, R.},
	month = dec,
	year = {2013},
	note = {ADS Bibcode: 2013MNRAS.436.2188R},
	pages = {2188--2231},
}

@article{safranek-shrader_star_2014,
	title = {Star formation in the first galaxies - {II}. {Clustered} star formation and the influence of metal line cooling},
	volume = {438},
	issn = {0035-8711},
	url = {https://ui.adsabs.harvard.edu/abs/2014MNRAS.438.1669S},
	doi = {10.1093/mnras/stt2307},
	urldate = {2022-12-06},
	journal = {\mnras},
	author = {Safranek-Shrader, Chalence and Milosavljević, Miloš and Bromm, Volker},
	month = feb,
	year = {2014},
	note = {ADS Bibcode: 2014MNRAS.438.1669S},
	pages = {1669--1685},
}

@article{schauer_influence_2019,
	title = {The influence of streaming velocities on the formation of the first stars},
	volume = {484},
	issn = {0035-8711},
	url = {https://ui.adsabs.harvard.edu/abs/2019MNRAS.484.3510S/abstract},
	doi = {10.1093/mnras/stz013},
	language = {en},
	number = {3},
	urldate = {2025-06-06},
	journal = {\mnras},
	author = {Schauer, Anna T. P. and Glover, Simon C. O. and Klessen, Ralf S. and Ceverino, Daniel},
	month = apr,
	year = {2019},
	pages = {3510},
}

@article{schauer_influence_2021,
	title = {The influence of streaming velocities and {Lyman}-{Werner} radiation on the formation of the first stars},
	volume = {507},
	issn = {0035-8711},
	url = {https://ui.adsabs.harvard.edu/abs/2021MNRAS.507.1775S/abstract},
	doi = {10.1093/mnras/stab1953},
	language = {en},
	number = {2},
	urldate = {2025-06-06},
	journal = {\mnras},
	author = {Schauer, Anna T. P. and Glover, Simon C. O. and Klessen, Ralf S. and Clark, Paul},
	month = oct,
	year = {2021},
	pages = {1775},
}

@article{schaerer_properties_2002,
	title = {On the properties of massive {Population} {III} stars and metal-free stellar populations},
	volume = {382},
	issn = {0004-6361},
	url = {https://ui.adsabs.harvard.edu/abs/2002A&A...382...28S},
	doi = {10.1051/0004-6361:20011619},
	urldate = {2025-03-11},
	journal = {\aap},
	author = {Schaerer, D.},
	month = jan,
	year = {2002},
	note = {ADS Bibcode: 2002A\&A...382...28S},
	pages = {28--42},
}

@article{schmidt_rate_1959,
	title = {The {Rate} of {Star} {Formation}.},
	volume = {129},
	issn = {0004-637X},
	url = {https://ui.adsabs.harvard.edu/abs/1959ApJ...129..243S/abstract},
	doi = {10.1086/146614},
	language = {en},
	urldate = {2026-01-22},
	journal = {\apj},
	author = {Schmidt, Maarten},
	month = mar,
	year = {1959},
	pages = {243},
}

@article{skinner_cradles_2020,
	title = {Cradles of the first stars: self-shielding, halo masses, and multiplicity},
	volume = {492},
	issn = {0035-8711},
	shorttitle = {Cradles of the first stars},
	url = {https://ui.adsabs.harvard.edu/abs/2020MNRAS.492.4386S/abstract},
	doi = {10.1093/mnras/staa139},
	language = {en},
	number = {3},
	urldate = {2026-05-30},
	journal = {\mnras},
	author = {Skinner, Danielle and Wise, John H.},
	month = mar,
	year = {2020},
	pages = {4386--4397},
}

@article{stacy_first_2010,
	title = {The first stars: formation of binaries and small multiple systems},
	volume = {403},
	issn = {0035-8711},
	shorttitle = {The first stars},
	url = {https://ui.adsabs.harvard.edu/abs/2010MNRAS.403...45S},
	doi = {10.1111/j.1365-2966.2009.16113.x},
	urldate = {2023-02-16},
	journal = {\mnras},
	author = {Stacy, Athena and Greif, Thomas H. and Bromm, Volker},
	month = mar,
	year = {2010},
	note = {ADS Bibcode: 2010MNRAS.403...45S},
	pages = {45--60},
}

@article{sugimura_birth_2020,
	title = {The {Birth} of a {Massive} {First}-star {Binary}},
	volume = {892},
	issn = {0004-637X},
	url = {https://ui.adsabs.harvard.edu/abs/2020ApJ...892L..14S},
	doi = {10.3847/2041-8213/ab7d37},
	urldate = {2022-07-14},
	journal = {\apj},
	author = {Sugimura, Kazuyuki and Matsumoto, Tomoaki and Hosokawa, Takashi and Hirano, Shingo and Omukai, Kazuyuki},
	month = mar,
	year = {2020},
	note = {ADS Bibcode: 2020ApJ...892L..14S},
	pages = {L14},
}

@article{sugimura_formation_2023,
	title = {Formation of {Massive} and {Wide} {First}-star {Binaries} in {Radiation} {Hydrodynamic} {Simulations}},
	volume = {959},
	issn = {0004-637X},
	url = {https://ui.adsabs.harvard.edu/abs/2023ApJ...959...17S},
	doi = {10.3847/1538-4357/ad02fc},
	urldate = {2023-12-11},
	journal = {\apj},
	author = {Sugimura, Kazuyuki and Matsumoto, Tomoaki and Hosokawa, Takashi and Hirano, Shingo and Omukai, Kazuyuki},
	month = dec,
	year = {2023},
	note = {ADS Bibcode: 2023ApJ...959...17S},
	pages = {17},
}

@article{sugimura_structure_2020,
	title = {Structure and instability of the ionization fronts around moving black holes},
	volume = {495},
	issn = {0035-8711},
	url = {https://ui.adsabs.harvard.edu/abs/2020MNRAS.495.2966S},
	doi = {10.1093/mnras/staa1394},
	urldate = {2024-01-20},
	journal = {\mnras},
	author = {Sugimura, Kazuyuki and Ricotti, Massimo},
	month = jul,
	year = {2020},
	note = {ADS Bibcode: 2020MNRAS.495.2966S},
	pages = {2966--2978},
}

@article{sugimura_violent_2024,
	title = {Violent {Starbursts} and {Quiescence} {Induced} by {Far}-ultraviolet {Radiation} {Feedback} in {Metal}-poor {Galaxies} at {High} {Redshift}},
	volume = {970},
	issn = {0004-637X},
	url = {https://ui.adsabs.harvard.edu/abs/2024ApJ...970...14S/abstract},
	doi = {10.3847/1538-4357/ad499a},
	language = {en},
	number = {1},
	urldate = {2025-08-07},
	journal = {\apj},
	author = {Sugimura, Kazuyuki and Ricotti, Massimo and Park, Jongwon and Garcia, Fred Angelo Batan and Yajima, Hidenobu},
	month = jul,
	year = {2024},
	pages = {14},
}

@article{susa_secondary_2006,
	title = {Secondary {Star} {Formation} in a {Population} {III} {Object}},
	volume = {645},
	issn = {0004-637X},
	url = {https://ui.adsabs.harvard.edu/abs/2006ApJ...645L..93S/abstract},
	doi = {10.1086/506275},
	language = {en},
	number = {2},
	urldate = {2025-09-09},
	journal = {\apj},
	author = {Susa, Hajime and Umemura, Masayuki},
	month = jul,
	year = {2006},
	pages = {L93--L96},
}

@article{susa_mass_2014,
	title = {The {Mass} {Spectrum} of the {First} {Stars}},
	volume = {792},
	issn = {0004-637X},
	url = {https://ui.adsabs.harvard.edu/abs/2014ApJ...792...32S},
	doi = {10.1088/0004-637X/792/1/32},
	urldate = {2022-09-22},
	journal = {\apj},
	author = {Susa, Hajime and Hasegawa, Kenji and Tominaga, Nozomu},
	month = sep,
	year = {2014},
	note = {ADS Bibcode: 2014ApJ...792...32S},
	pages = {32},
}

@article{tegmark_how_1997,
	title = {How {Small} {Were} the {First} {Cosmological} {Objects}?},
	volume = {474},
	issn = {0004-637X},
	url = {https://ui.adsabs.harvard.edu/abs/1997ApJ...474....1T},
	doi = {10.1086/303434},
	urldate = {2023-06-06},
	journal = {\apj},
	author = {Tegmark, Max and Silk, Joseph and Rees, Martin J. and Blanchard, Alain and Abel, Tom and Palla, Francesco},
	month = jan,
	year = {1997},
	note = {ADS Bibcode: 1997ApJ...474....1T},
	pages = {1},
}

@article{teyssier_cosmological_2002,
	title = {Cosmological hydrodynamics with adaptive mesh refinement. {A} new high resolution code called {RAMSES}},
	volume = {385},
	issn = {0004-6361},
	url = {https://ui.adsabs.harvard.edu/abs/2002A&A...385..337T},
	doi = {10.1051/0004-6361:20011817},
	urldate = {2023-09-07},
	journal = {\aap},
	author = {Teyssier, R.},
	month = apr,
	year = {2002},
	note = {ADS Bibcode: 2002A\&A...385..337T},
	pages = {337--364},
}

@article{trenti_formation_2009,
	title = {Formation {Rates} of {Population} {III} {Stars} and {Chemical} {Enrichment} of {Halos} during the {Reionization} {Era}},
	volume = {694},
	issn = {0004-637X},
	url = {https://ui.adsabs.harvard.edu/abs/2009ApJ...694..879T/abstract},
	doi = {10.1088/0004-637X/694/2/879},
	language = {en},
	number = {2},
	urldate = {2025-08-15},
	journal = {\apj},
	author = {Trenti, Michele and Stiavelli, Massimo},
	month = apr,
	year = {2009},
	pages = {879},
}

@article{turk_formation_2009,
	title = {The {Formation} of {Population} {III} {Binaries} from {Cosmological} {Initial} {Conditions}},
	volume = {325},
	issn = {0036-8075},
	url = {https://ui.adsabs.harvard.edu/abs/2009Sci...325..601T},
	doi = {10.1126/science.1173540},
	urldate = {2023-06-27},
	journal = {Science},
	author = {Turk, Matthew J. and Abel, Tom and O'Shea, Brian},
	month = jul,
	year = {2009},
	note = {ADS Bibcode: 2009Sci...325..601T},
	pages = {601},
}

@article{whalen_finding_2014,
	title = {Finding the {First} {Cosmic} {Explosions}. {III}. {Pulsational} {Pair}-instability {Supernovae}},
	volume = {781},
	issn = {0004-637X},
	url = {https://ui.adsabs.harvard.edu/abs/2014ApJ...781..106W/abstract},
	doi = {10.1088/0004-637X/781/2/106},
	language = {en},
	number = {2},
	urldate = {2025-05-14},
	journal = {\apj},
	author = {Whalen, Daniel J. and Smidt, Joseph and Even, Wesley and Woosley, S. E. and Heger, Alexander and Stiavelli, Massimo and Fryer, Chris L.},
	month = feb,
	year = {2014},
	pages = {106},
}

@article{wells_connecting_2022,
	title = {Connecting {Primordial} {Star}-forming {Regions} and {Second}-generation {Star} {Formation} in the {Phoenix} {Simulations}},
	volume = {932},
	issn = {0004-637X},
	url = {https://ui.adsabs.harvard.edu/abs/2022ApJ...932...71W/abstract},
	doi = {10.3847/1538-4357/ac6c87},
	language = {en},
	number = {1},
	urldate = {2026-06-02},
	journal = {\apj},
	author = {Wells, Azton I. and Norman, Michael L.},
	month = jun,
	year = {2022},
	pages = {71},
}

@article{wise_resolving_2008,
	title = {Resolving the {Formation} of {Protogalaxies}. {III}. {Feedback} from the {First} {Stars}},
	volume = {685},
	issn = {0004-637X},
	url = {https://ui.adsabs.harvard.edu/abs/2008ApJ...685...40W},
	doi = {10.1086/590417},
	urldate = {2022-12-06},
	journal = {\apj},
	author = {Wise, John H. and Abel, Tom},
	month = sep,
	year = {2008},
	note = {ADS Bibcode: 2008ApJ...685...40W},
	pages = {40--56},
}

@article{wise_birth_2012,
	title = {The {Birth} of a {Galaxy}: {Primordial} {Metal} {Enrichment} and {Stellar} {Populations}},
	volume = {745},
	issn = {0004-637X},
	shorttitle = {The {Birth} of a {Galaxy}},
	url = {https://ui.adsabs.harvard.edu/abs/2012ApJ...745...50W},
	doi = {10.1088/0004-637X/745/1/50},
	urldate = {2022-09-24},
	journal = {\apj},
	author = {Wise, John H. and Turk, Matthew J. and Norman, Michael L. and Abel, Tom},
	month = jan,
	year = {2012},
	note = {ADS Bibcode: 2012ApJ...745...50W},
	pages = {50},
}

@article{xu_heating_2014,
	title = {Heating the {Intergalactic} {Medium} by {X}-{Rays} from {Population} {III} {Binaries} in {High}-redshift {Galaxies}},
	volume = {791},
	issn = {0004-637X},
	url = {https://ui.adsabs.harvard.edu/abs/2014ApJ...791..110X},
	doi = {10.1088/0004-637X/791/2/110},
	urldate = {2022-07-06},
	journal = {\apj},
	author = {Xu, Hao and Ahn, Kyungjin and Wise, John H. and Norman, Michael L. and O'Shea, Brian W.},
	month = aug,
	year = {2014},
	note = {ADS Bibcode: 2014ApJ...791..110X},
	pages = {110},
}

@article{xu_x-ray_2016,
	title = {X-{Ray} {Background} at {High} {Redshifts} from {Pop} {III} {Remnants}: {Results} from {Pop} {III} {Star} {Formation} {Rates} in the {Renaissance} {Simulations}},
	volume = {832},
	issn = {0004-637X},
	shorttitle = {X-{Ray} {Background} at {High} {Redshifts} from {Pop} {III} {Remnants}},
	url = {https://ui.adsabs.harvard.edu/abs/2016ApJ...832L...5X},
	doi = {10.3847/2041-8205/832/1/L5},
	urldate = {2022-07-06},
	journal = {\apj},
	author = {Xu, Hao and Ahn, Kyungjin and Norman, Michael L. and Wise, John H. and O'Shea, Brian W.},
	month = nov,
	year = {2016},
	note = {ADS Bibcode: 2016ApJ...832L...5X},
	pages = {L5},
}

@article{yoshida_early_2007,
	title = {Early {Cosmological} {H} {II}/{He} {III} {Regions} and {Their} {Impact} on {Second}-{Generation} {Star} {Formation}},
	volume = {663},
	issn = {0004-637X},
	url = {https://ui.adsabs.harvard.edu/abs/2007ApJ...663..687Y},
	doi = {10.1086/518227},
	urldate = {2022-07-07},
	journal = {\apj},
	author = {Yoshida, Naoki and Oh, S. Peng and Kitayama, Tetsu and Hernquist, Lars},
	month = jul,
	year = {2007},
	note = {ADS Bibcode: 2007ApJ...663..687Y},
	pages = {687--707},
}




\appendix

\section{Volume~2 as the Representative Volume}
\label{sec:vol2}

Estimating the X-ray background requires the cosmic star formation history (SFH) to approximate the emissivity at earlier epochs. For simplicity, we use the SFH of Volume~2 as a proxy for the cosmic mean, since its halo mass function closely matches the cosmic average at $z = 9$. Here we assess the impact of this choice.

\begin{figure}
    \centering
	\includegraphics[width=0.48\textwidth]{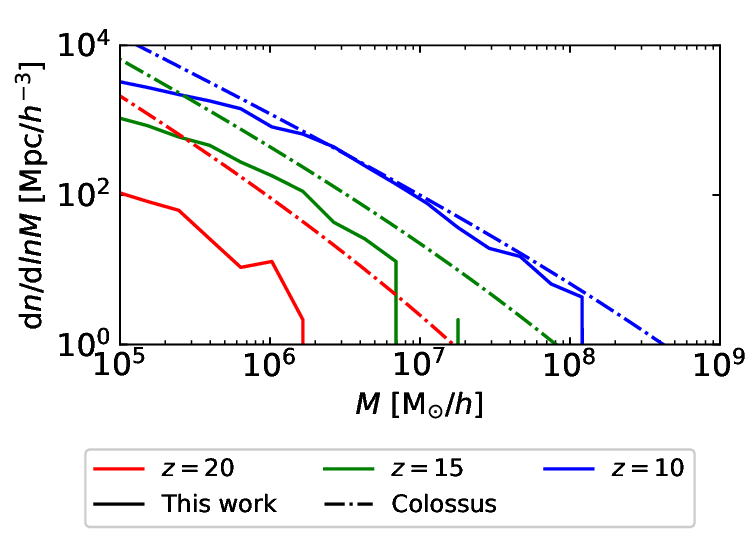}
    \caption{Halo mass functions at $z = 10, 15,$ and 20 shown in blue, green, and red, respectively. The results from Volume~2 (solid lines) are compared to the Press-Schechter (dotted dashed lines) computed using \colossus \citep{diemer_colossus_2018}.}
    \label{fig:compare_mf}
\end{figure}

Fig.~\ref{fig:compare_mf} compares the halo mass functions of Volume~2 (solid lines) with the Press-Schechter (PS) fits (dot-dashed lines) at $z = 10$, $15$, and $20$ (blue, green, and red). At $z = 10$, the two agree well within the valid halo mass range ($10^6~\msun \lesssim M_{halo} \lesssim 10^8~\msun$), justifying our choice. At higher redshifts, however, the Volume~2 halo abundance falls below the PS prediction, with the discrepancy growing towards $z = 20$.

We expect this deviation to have only a minor impact. As discussed in Section~2.7 of PR26a (equations~14 and 17), the emissivity estimate depends on $\npop$. At high redshifts, Pop~III star formation is stochastic due to the small number of forming stars, introducing uncertainty in the X-ray intensity. To reduce this uncertainty, we estimate $\npop$ at $z > z_{ana} \approx 18$ using the PS fit with a fixed critical mass of $M_{crit,1} = 2 \times 10^6~\msun$, so the underestimated halo mass function has little influence above $z_{ana}$.

Although the transition at $z_{ana}$ (equation~(17) of PR26a) may cause a slight underestimate of $\npop$, the X-ray intensity evolves smoothly across this redshift, reflecting the cumulative SFH rather than the instantaneous rate (fig.~7 of PR26a).

\section{Effects of the Strength of the X-ray Background}
\label{sec:xraytest}

\begin{figure}
    \centering
	\includegraphics[width=0.48\textwidth]{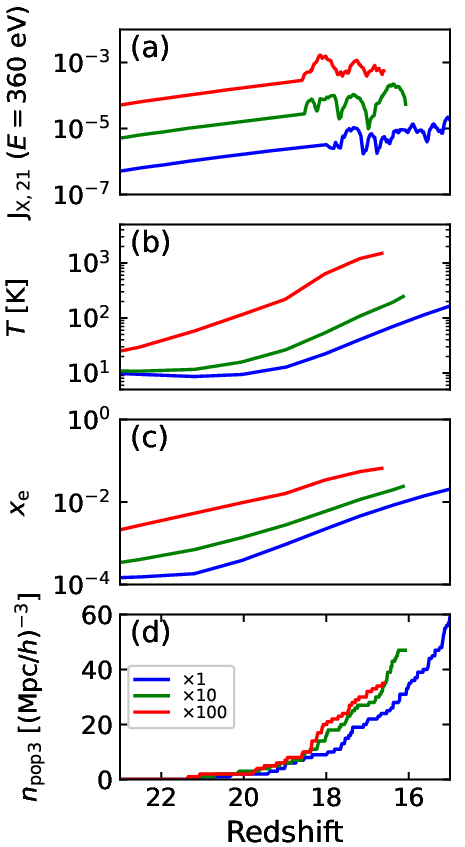}
    \caption{\textbf{Panel~a:} X-ray intensity ($E = 360$~eV) as a function of redshift for fiducial (blue) and enhanced (green and red) X-ray emissivity (see equation~(16) of PR26a). The intensity scales approximately linearly with the emissivity. \textbf{Panel~b:} Volume-weighted mean gas temperature. The difference between the fiducial and $\times10$ runs remains below a factor of 10 due to gas cooling. \textbf{Panel~c:} Volume-weighted mean electron fraction, which follows a similar trend to the gas temperature. The small difference between the first two cases arises from recombination. \textbf{Panel~d:} Number of Pop~III stars. Pop~III star formation is promoted when the X-ray emissivity increases by a factor of 10, but no further increase occurs at higher value ($\times 100$) due to X-ray heating.}
    \label{fig:xray1}
\end{figure}

In our fiducial model, Pop~III supernovae produce a weak ($\intj_{21} \sim 10^{-5}$) X-ray background that yields only mild positive feedback. If additional sources such as AGNs or HMXBs are present, the feedback could be stronger. Although a self-consistent treatment of these sources is left for future work, we perform test simulations with artificially enhanced X-ray backgrounds by a factor 10 and 100 to assess whether stronger X-rays can further promote Pop~III star formation. We note that these sources predominantly emit hard X-rays ($E \sim 2.0$--$10.0$~keV), so their effects cannot be directly compared to the enhanced soft X-ray ($E \sim 0.2$--$2.0$~keV) backgrounds adopted here.

Fig.~\ref{fig:xray1} presents the results. We increase the X-ray emissivity by factors of 10 (green) and 100 (red) relative to the fiducial case (blue; see equation~(16) of PR26a). Panel~a shows that the X-ray intensity at $E = 360$~eV scales with the emissivity. The earlier onset of fluctuations in the enhanced runs reflects the earlier crossing of the Pop~III star count threshold used to estimate the background (i.e., $z_{ana}$).

Panels~b and c show X-ray heating and ionisation. The volume-weighted mean temperature increases only weakly with X-ray intensity — at most a factor of $\approx 2$ for $\times 10$ — as gas cooling counteracts X-ray heating. Similarly, the electron fraction does not scale linearly with X-ray intensity due to recombination.

Panel~d shows that the increase in $\npop$ is much smaller than the increase in X-ray intensity. We interpret this as a consequence of strong X-ray heating suppressing Pop~III star formation in low-mass haloes ($T_{vir} \sim 10^3$~K), which limits the net positive effect of the enhanced electron fraction.

\section{Estimation of $L_{X}$-SFR Relation}
\label{sec:Lx}

In this section, we describe how we estimate the X-ray luminosities ($L_{X}$) of galaxies for different X-ray sources discussed in Section~\ref{sec:obs}.

\textbf{Pop~III SFR: } Although Pop~III star formation in a minihalo is a single event, we estimate its SFR per comoving (Mpc/$h$)$^3$ to compare the relative contributions of Pop~III and Pop~II stars to the X-ray emissivity. Panel~d of Fig.~\ref{fig:npop3} shows that Pop~III SFR is $\sim 10^{-6}$~yrs or $\sim 10^{-4}~\msun$~yr$^{-1}$ on average in  the target region of Volume~1 and Volume~2 with box sizes of 1 Mpc/$h$. 

\textbf{Pop~II SFR: } We identified all haloes in the final snapshots and counted Pop~II star particles within their virial radii. Our analysis focus on Pop~II haloes in Volumes~1 and 2, while haloes in the other volumes are excluded because they either lie outside the target region (i.e., susceptible to numerical artefacts) or have only a few particles. The Pop~II SFR is measured over time intervals of $\approx 5$~Myr for each galaxy. We use the peak SFRs and stellar masses at those epochs in the analysis.

\textbf{X-ray Luminosity:} We define the X-ray luminosity as $L_{X} = (f_{X}/t_{*})E_{X}$ where $f_{X}$ is the fraction of stars that become X-ray sources (SNe or XRBs), $t_{*}$ is the characteristic star formation timescale, and $E_{X}$ is the total X-ray energy. The factor $f_{X}/t_{*}$ defines the rate of formation of X-ray sources. The star formation time scale $t_{*}$ is,
\begin{equation}
    t_{*} \sim 1~\mathrm{Myr}~\left( \frac{M}{100~\msun} \right) \left( \frac{10^{-4}~\msun/{\mathrm{yr}}}{\mathrm{SFR}} \right),
    \label{eq:tsf}
\end{equation}
where $M$ is the stellar mass of an object in $\msun$, and SFR is the star formation rate in $\msun$~yr$^{-1}$. The X-ray luminosity is, therefore,
\begin{equation}
    L_{X} = 
    \left( \frac{f_{X}E_{X}}{1~{\mathrm{Myr}}} \right)
    \left( \frac{100~\msun}{M} \right) \left( \frac{\mathrm{SFR}}{10^{-4}~\msun/\mathrm{yr}} \right).
\end{equation}

\textbf{Pop~III Supernovae:} The X-ray energy is $E_{X} = 6 \times 10^{50}$~erg (100 times more energetic than a normal SN, see PR26a) and $t_{*} \sim 1$~Myr (see \textbf{Pop~III SFR}). We also assume that all Pop~III stars explode as PISNe and only a single objects forms per halo, that is, $f_{PISN} = 1$. Therefore, the X-ray luminosity of Pop~III minihaloes per comoving (Mpc/$h$)$^3$ is,
\begin{equation}
    \frac{L_{X}}{V} \sim \frac{6 \times 10^{50}~\mathrm{erg}}{1~\mathrm{Myr}~(\mathrm{Mpc}/h)^{3}} \sim 2 \times 10^{37}~\mathrm{erg~s^{-1}(Mpc/\textit{h})^{-3}}.
    \label{eq:sn3}
\end{equation}
For a reference, \citet{whalen_finding_2014} predicted that the bolometric luminosity of a PISN remains $\sim 10^{43}$~erg~s$^{-1}$ for the first $\sim 10^3$~yr. We highlight that here we estimated an average X-ray luminosity per comoving (Mpc/$h$)$^3$ from a Pop~III supernovae to compare this to Pop~II SNe. Pop~III supernovae are less luminous than Pop~II X-ray sources discussed below. Consequently, the dominant X-ray sources may transition from Pop~III to Pop~II following the transition of the star formation mode.

\textbf{Pop~III HMXBs:} In estimating the X-ray luminosity from Pop~II XRBs, we adopt the assumptions used by previous studies \citep{jeon_radiative_2014, hummel_first_2015}: binaries accrete gas at the Eddington rate ($\sim 10^{38}~$erg~s$^{-1}$ per $1~\msun$), $30~\%$ of the released energy is emitted in X-rays, and the accretion phase lasts for $\sim 2$ Myr. We further assume that the typical mass of Pop~III HMXBs is $30~\msun$. The total X-ray energy emitted by a ``single'' HMXB is $\sim 6 \times 10^{52}$ erg. Regarding the HMXB formation timescale, we assumed 1/30 of Pop~III binaries survive as Pop~III HMXBs (i.e., $f_{HMXB} = 1/30$, binary formation timescale is 30 times longer) following the aforementioned studies. The X-ray emissivity is,
\begin{equation}
    \frac{L_{X}}{V} \sim \frac{1}{30} \frac{6 \times 10^{52}~\mathrm{erg}}{1~\mathrm{Myr}~(\mathrm{Mpc}/h)^{3}} \sim 6 \times 10^{37}~\mathrm{erg~s^{-1} (Mpc/\textit{h})^{-3}}.
    \label{eq:xrb3}
\end{equation}

\textbf{Pop~II Supernovae:} For Pop~II SN feedback, we adopt the standard thermal feedback model implemented in \ramses \citep{dubois_agn-driven_2013}. In this model, a SN occurs every $10~\msun$ and therefore the total SN energy emitted by a Pop~II star particle of mass $M$ is $10^{51}(M/10~\msun)$~erg. We assume the X-ray energy is proportional to the total SN energy, such that a galaxy with stellar mass $M_{gal}$ emits $E_{X} = 6 \times 10^{48}~(M_{gal}/10~\msun)$~erg of energy in X-rays (see R16 and PR26a for numerics). We further assume that one in hundred stars explode as a core-collapse SN (i.e., $f_{X} = 0.01$\footnote{If we adopt an IMF with a slope $\alpha = 2.3$, this assumption is consistent with the standard thermal SN feedback of \ramses aforementioned which assumes one SN occurs every $10~\msun$.}) following the value reported by \citet{power_primordial_2009} with Kroupa IMF \citep{kroupa_variation_2001}. The X-ray luminosity of a galaxy is given by
\begin{equation}
    L_{X} \sim 2 \times 10^{37} \left( \frac{\mathrm{SFR} }{0.1~ \msun~\mathrm{yr}^{-1}}\right)~\mathrm{erg~s^{-1}}.
    \label{eq:sn2}
\end{equation}
Note that both $t_{*}$ and $E_{X}$ are proportional to the mass of the galaxy, $M$, and therefore cancel in the above equation. As a result, $L_{X}$ depends only on the SFR. For example, a galaxy with the $M_{gal} = 10^6~\msun$ and SFR$=10^{-1}~\msun$~yr$^{-1}$ have $E_{X} \sim 6 \times 10^{53}$~erg and $t_{*} \sim 10$~Myr, and thereby emits $L_{X} = 2 \times 10^{37}$~erg~s$^{-1}$.

\begin{figure}
    \centering
	\includegraphics[width=0.48\textwidth]{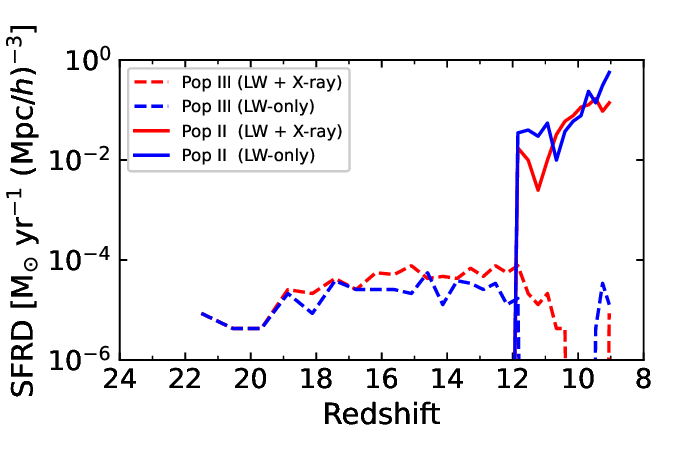}
    \caption{Star formation rate densities of Volume~2 in the absence (blue) and presence (red) of the X-ray background. Dashed and solid lines represent SFRDs of Pop~III and Pop~II stars, respectively. At $z \gtrsim 12$, Pop~III SFRDs remain $\sim 10^{-4}~\msun$~yr$^{-1}$ in both runs and are consistent with the values reported by previous theoretical studies \citep{skinner_cradles_2020,wells_connecting_2022,hegde_self-consistent_2023,hegde_efficient_2025}. Pop~II SFRDs at $z \sim 12$--$9$ is $\sim 10^{-2}$--$10^{0}~\msun$~yr$^{-1}$~(Mpc$/h)^{-3}$.}
    \label{fig:sfrd}
\end{figure}

\textbf{Pop~II X-ray Binaries (XRBs):} We adopted the same approach to Pop~III HMXHBs but with different binary mass ($5~\msun$). Under these assumptions, a ``single'' XRB emits a total X-ray energy of $\sim 3 \times 10^{51}$ erg. We adopt a binary fraction of $0.3$ reported by \citep{power_primordial_2009} and further assume that all massive stars ($M \gtrsim 10~\msun$) are in binaries. A single Pop~II particle therefore emits $0.3 \times 10^{52} (M/10~\msun)$~erg in X-rays. $f_{X}$ is given by the product of the binary fraction (0.3) and the SN fraction (0.01), yielding $f_{X} = 0.003$. The X-ray luminosity of a galaxy  is therefore,
\begin{equation}
    L_{X} \sim 3 \times 10^{39} \left( \frac{\mathrm{SFR} }{0.1~ \msun~\mathrm{yr}^{-1}}\right)~\mathrm{erg~s^{-1}}.
    \label{eq:xrb2}
\end{equation}
The Pop~II SFR per comoving (Mpc/$h$)$^3$ at $z\sim 10-12$ is roughly $0.1$~M$_\odot$~yr$^{-1}$~(Mpc/$h$)$^{-3}$ (Fig.~\ref{fig:sfrd}), therefore the X-ray emissivity from Pop~II stars XRBs is $3 \times 10^{39}$~erg~s$^{-1}$~(Mpc/$h$)$^3$ that is about 50 times higher than our estimate from Pop~III stars. Instead, using Eq.~\ref{eq:sn2}, the X-ray emissivity from SNe from Pop~II stars is comparable to the one from Pop~III stars.

\bsp	
\label{lastpage}
\end{document}